# Decentralized Frequency Regulation of Hybrid MTDC-linked Grids

Young-Jin Kim

***Abstract*—This paper proposes a new strategy for optimal grid frequency regulation (FR) in an interconnected power system where regional ac grids and an offshore wind farm are linked via a multi-terminal high voltage direct-current (MTDC) network. In the proposed strategy, decentralized $H_\infty$ controllers are developed to coordinate the operations of ac synchronous generators and hybrid MTDC converters, thus achieving optimal power sharing of interconnected ac grids and minimizing frequency deviations in each grid. To develop the controllers, robust optimization problems are formulated and solved using a dynamic model of the hybrid MTDC-linked grids with model parameter uncertainty and decentralized control inputs and outputs. The model orders of the resulting controllers are then reduced using a balanced truncation algorithm to eliminate unobservable and uncontrollable state variables while preserving their dominant response characteristics. Sensitivity and eigenvalue analyses are conducted focusing on the effects of grid measurements, parameter uncertainty levels, and communication time delays. Comparative case studies are also carried out to verify that the proposed strategy improves the effectiveness, stability, and robustness of real-time FR in MTDC- linked grids under various conditions characterized mainly by load demands, communications systems, and weighting functions.**

## I. Introduction

SUSTAINABLE wind energy is critical to reducing fossil fuel emissions and mitigating global climate change [1]. In particular, the capacity of offshore wind farms (OWFs) is growing faster for many reasons, including a lack of suitable onshore sites and better wind conditions of offshore sites [2]. Meanwhile, high-voltage direct-current (HVDC) technology has been increasingly considered a promising solution for delivering offshore wind power to onshore power grids and sharing it between interconnected regional grids [3], [4].

In HVDC technology, line-commutated converter (LCC) and voltage source converter (VSC) have been widely used due to their distinct characteristics. An LCC is characterized by, for example, a high capacity for bulk power transmission, low power loss, and low installation cost [5]. However, it operates only as a rectifier or an inverter that can control either dc voltage or dc current [6]. By contrast, a VSC can change the direction of dc current and independently control active and reactive power, without the need for reactive power compensators, although it has lower power rating and higher power loss even when implemented in the form of modular multilevel converter [7]. In several projects, there is a recent interest in combining the advantages of both LCC and VSC, developing hybrid schemes [5].

An HVDC system can be further developed in the form of a multi-terminal dc (MTDC) system [9]–[14], for example, where LCCs, grid-side VSCs (GSVSCs), and OWF-side VSCs (WFVSCs) are linked via a dc network to connect regional ac grids with OWFs. This enhances the flexibility of inter-grid power delivery and sharing, thereby facilitating the integration of large-scale OWFs with regional load centers. However, due to the interconnection, each grid frequency is affected by power imbalances not only in the corresponding grid but also in other grids [8], implying that intermittent wind power causes frequency deviations in all interconnected grids. This motivates the development of a proper strategy for power balancing in MTDC-linked grids, essentially requiring the coordinated control of regional power generation and ac-to-dc power transmission.

Several studies have been conducted on frequency regulation (FR) in MTDC-linked grids. For example, in [9], VSCs were droop-controlled to mitigate grid frequency deviations and dc voltage variations resulting from wind power generation. In [10], communication-free coordination of VSC terminals was accomplished using a $V_{dc}$-$f$ droop control scheme. In [11], adaptive droop control of a VSC-MTDC system was discussed, wherein the $V_{dc}$-$I_{dc}$-$f$ characteristics of the system were analyzed to determine the droop control gains. However, in [9]–[11], FR relied mainly on droop control or, equivalently, primary frequency control, rather than optimal secondary frequency control.

The optimal FR of MTDC-linked grids has been discussed in recent studies (e.g., [12]–[14]). In [12], a model predictive control (MPC) algorithm was adopted for the optimal power sharing in a VSC-MTDC system, wherein a discrete-time optimization problem was formulated considering grid operating costs. In [13] and [14], the MPC-based power sharing was also achieved to minimize the frequency deviations in the MTDC-linked grids with only VSCs and LCCs, respectively. However, in [12]–[14], hybrid MTDC systems were not considered. Moreover, in general, MPC is computationally intensive, because an optimization problem for MPC needs to be solved directly at every time step. MPC is also likely to cause oscillatory and even unstable operation of MTDC-linked grids due to communication time delays and model parameter uncertainty [15]. Therefore, the optimal FR strategies discussed in [12]–[14] need to be further analyzed under practical conditions of MTDC systems, for example, with regard to hybrid converters, communications links, and system parameter estimates.

When system parameters are uncertain, robust controllers can effectively reduce the effect of disturbance on system operation. For example, in [16], the robust dc voltage control of a VSC was achieved by establishing an $H_\infty$-mixed sensitivity function in a linear matrix inequality framework, wherein dc current injection

This research was supported by the National Research Foundation of Korea (NRF) grant funded by the Korea government (*corresponding author: Y. Kim*).
Y. Kim is with the Department of Electrical Engineering, Pohang University of Science and Technology (POSTECH), Pohang 790–784, South Korea (e-mail: powersys@postech.ac.kr).

was set as a disturbance input. In [17], a systematic method was provided to design the $H_\infty$ controller of a VSC for optimal and robust control of active power and ac voltage. In [18] and [19], $H_2$ and $H_\infty$ controllers were discussed to optimally regulate the dc-link voltage and the converter internal energy of an HVDC system, respectively. However, in [16]–[19], robust control was applied to only a single converter; the coordination with synchronous generators and the power sharing between regional ac grids were not considered. In [19], a centralized optimization problem was formulated for centralized and decentralized $H_\infty$ control of the dc-link voltage in a modular multilevel converter when applied to a VSC-HVDC system. The centralized controller slightly outperformed the decentralized controller. However, in practice, such a centralized controller is difficult to implement and apply for MTDC-linked grids due to computational and communication overheads.

This paper proposes a new decentralized strategy for optimal robust FR in regional ac grids that are interconnected via a hybrid MTDC system including LCCs and VSCs. For each ac grid, a decentralized $H_\infty$ controller is developed for coordinated control of generated and transmitted power, optimizing inter-grid power sharing and hence minimizing frequency deviations. Specifically, for the proposed strategy, a dynamic model of hybrid MTDC-linked grids is implemented with decentralized control inputs and outputs. Given the dynamic model, a robust optimization problem for the optimal FR in each grid is formulated considering model parameter uncertainty. The solution to the problem leads to the optimal gain of the decentralized $H_\infty$ controller. A balanced truncation algorithm is then applied to reduce the controller model order, thus facilitating its practical implementation and the reference signal generation. Sensitivity and eigenvalue analyses are conducted to verify grid frequency stability in the proposed strategy, focusing on the effects of system parameters, parameter uncertainty, and communication time delay. Comparative case studies are also carried out to verify that the proposed strategy improves the effectiveness and robustness in reducing frequency deviations under various grid conditions.

The main contributions of this paper are summarized below:

- To the best of our knowledge, this is the first study reporting the optimal robust FR of MTDC-linked grids using decentralized $H_\infty$ controllers. Each $H_\infty$ controller is designed with reduced requirement on inter-grid measurements and communications.
- Hybrid MTDC-linked grids are modeled with decentralized control inputs and outputs. For each grid, considering model parameter uncertainty, an optimization problem is then formulated to determine the corresponding optimal robust control gains.
- The model orders of the optimal decentralized $H_\infty$ controllers are reduced using a balanced truncation algorithm, facilitating practical applications of the proposed FR strategy.
- The proposed strategy is comprehensively evaluated via comparison to conventional decentralized PI and $H_\infty$ controller-based strategies with regard to effectiveness and robustness against time-varying power imbalance, communication failures and time delays, and different weighting function models.

## II. Hybrid MTDC-linked Grids

Fig. 1 shows a schematic diagram of the proposed decentralized strategy for the optimal robust FR in MTDC-linked grids, where regional ac grids and OWFs are interconnected with each other using hybrid MTDC converters (i.e., VSCs and LCCs). Each grid includes multiple synchronous generators (SGs), loads, and ac transmission lines, and each OWF contains several wind turbine generators (WTGs) using permanent magnet synchronous generators (PMSGs). Note that for brevity, Fig. 1 illustrates an SG and a WTG. Moreover, for each WTG, a back-to-back converter and its maximum-power-point-tracking controller are modeled using simple equations for power balance and duty ratio update [35], because they marginally affect the regional frequency deviations. For the optimal FR in each grid, a decentralized $H_\infty$ controller receives measurements from the corresponding grid and interconnected grids via communication systems with time delays. The decentralized $H_\infty$ controller then generates optimal reference signals for the local controllers of the SGs and MTDC converter to adjust the generated and transmitted power in a coordinated manner. The references of ac voltage variations are set to zero, while focusing on the FR.

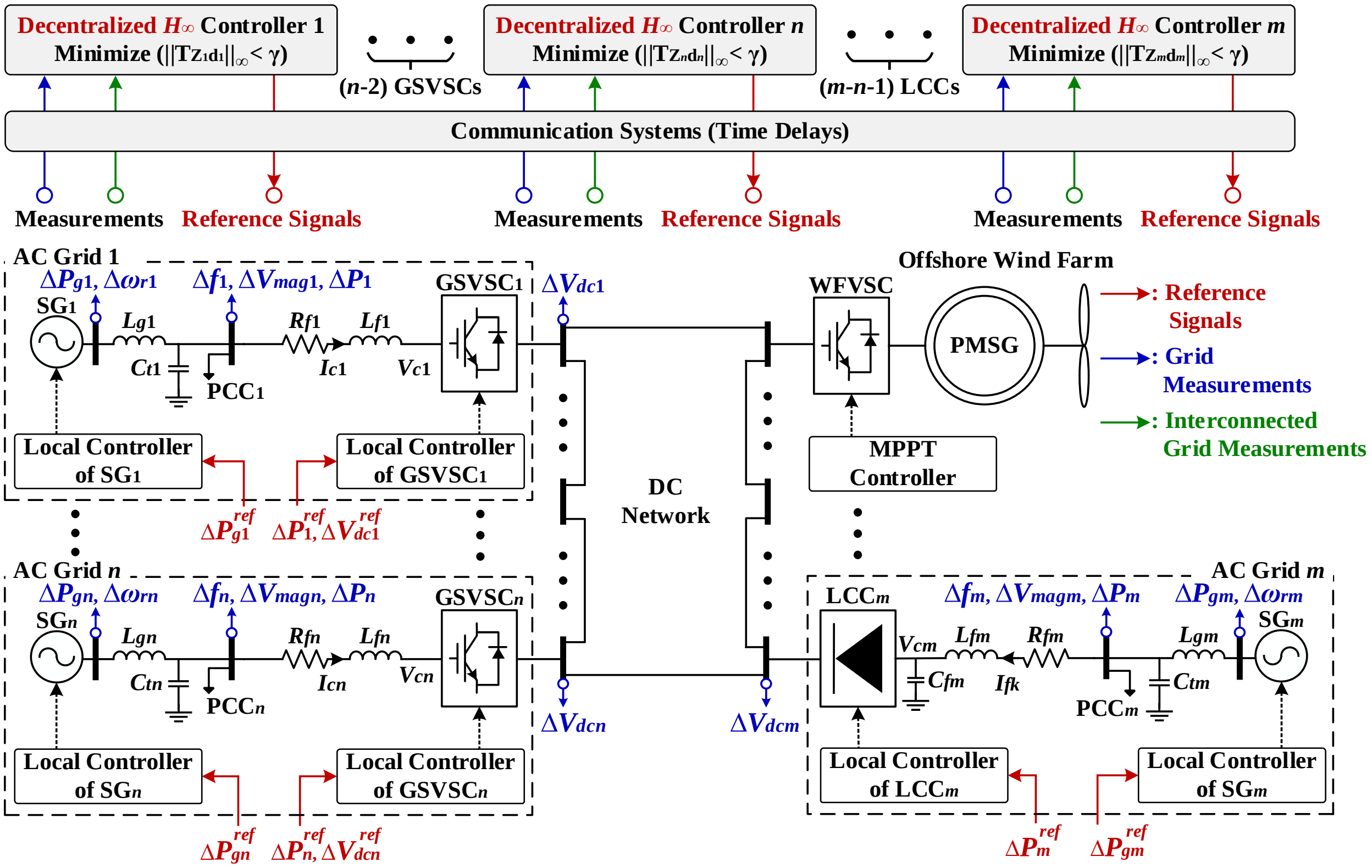

Fig. 1. A schematic diagram of the proposed decentralized strategy for optimal FR in MTDC-linked grids.

### A. *Modeling of Hybrid MTDC-linked Grids*

Without loss of generality, this paper considers three regional ac grids and one PMSG-type OWF, all of which are linked to a dc network via two GSVSCs, one LCC, and one WFVSC. Specifically, Grids 1 and 2 are interfaced using GSVSCs and Grid 3 is connected with the LCC at the points of common coupling (PCCs): i.e., $n = 2$ and $m = 3$ in Fig. 1. The GSVSCs are equipped with local $P$-$V_{dc}$ droop controllers, so that power transmitted from Grids 1 and 2 are instantaneously adjusted according to variations in the corresponding dc terminal voltages. On the other hand, the LCC operates as a rectifier with a local controller that transmits constant or time-varying power from the regional grid to the dc network. Note that the modeling and control procedures, discussed in Sections II and III, can readily be applied to a general case with more than three regional grids and different types of OWFs.

A dynamic model of the hybrid MTDC-linked grids is implemented by integrating the individual models of the ac grids, OWF, and dc network [see (A11)–(A16) and (B1)–(B5) in Appendix], as:

$$\dot{\mathbf{X}} = \mathbf{A_C}\cdot\mathbf{X} + \mathbf{B_r}\cdot\mathbf{r} + \mathbf{B_w}\cdot\mathbf{w}_d + \mathbf{E_C}\cdot\mathbf{Z}_{dc}\,, \tag{1}$$

$$\mathbf{Y} = \mathbf{C}\cdot\mathbf{X}\,, \tag{2}$$

where $\mathbf{A_C} = diag(\mathbf{A}_1,\ \mathbf{A}_2,\ \mathbf{A}_3,\ \mathbf{A}_{owf},\ \mathbf{A}_{dc})\,,$ (3)

$$\mathbf{B_r} = \begin{bmatrix} \mathbf{B}_{r1} & & \\ & \mathbf{B}_{r2} & \\ & & \mathbf{B}_{r3} \\ \mathbf{O} & \mathbf{O} & \mathbf{O} \\ \mathbf{O} & \mathbf{O} & \mathbf{O} \end{bmatrix}, \quad \mathbf{B_w} = \begin{bmatrix} \mathbf{B}_{w1} & & & \\ & \mathbf{B}_{w2} & & \\ & & \mathbf{B}_{w3} & \\ & & & \mathbf{B}_{owf} \\ \mathbf{O} & \mathbf{O} & \mathbf{O} & \mathbf{O} \end{bmatrix}, \tag{4}$$

$$\mathbf{C} = \begin{bmatrix} \mathbf{C}_1 & & & \mathbf{O} & \mathbf{O} \\ & \mathbf{C}_2 & & \mathbf{O} & \mathbf{O} \\ & & \mathbf{C}_3 & \mathbf{O} & \mathbf{O} \end{bmatrix},\ \mathbf{E_C} = diag(\mathbf{E}_1,\ \mathbf{E}_2,\ \mathbf{E}_3, \mathbf{E}_{owf}, \mathbf{E}_{dcv})\,, \tag{5}$$

$$\mathbf{X} = [\mathbf{X}_1^T,\ \mathbf{X}_2^T,\ \mathbf{X}_3^T,\ \mathbf{X}_{owf}^T, \mathbf{X}_{dc}^T]^T,\quad \mathbf{Y} = [\mathbf{Y}_1^T,\ \mathbf{Y}_2^T,\ \mathbf{Y}_3^T]^T\,, \tag{6}$$

$$\mathbf{r} = [\mathbf{r}_1^T,\ \mathbf{r}_2^T,\ \mathbf{r}_3^T]^T\,,\quad \mathbf{w}_d = [\Delta P_{L1},\ \Delta P_{L2},\ \Delta P_{L3},\ \Delta V_W]^T\,, \tag{7}$$

$$\mathbf{Z}_{dc} = [\mathbf{I}_{dc1},\ \mathbf{I}_{dc2},\ \mathbf{I}_{dc3},\ \mathbf{I}_{dc4},\ \mathbf{V}_{dc}]^T\,. \tag{8}$$

The detailed procedure for (1)–(8) is provided in Appendix. In (6), $\mathbf{X}_k$ includes the frequency $f_k$ and dc terminal voltage $V_{dck}$ for each grid $k$. Moreover, $\mathbf{X}_{dc}$ includes the dc currents $\mathbf{I}_{dc1\text{–}4}$ [see (8)] that flow through the dc transmission lines, as:

$$\mathbf{I}_{dc1-4} = \begin{bmatrix} 1 & 1 & 0 & 0 \\ -1 & 0 & 1 & 0 \\ 0 & -1 & 0 & 1 \\ 0 & 0 & -1 & -1 \end{bmatrix}\cdot\mathbf{X}_{dc} = \mathbf{F}_{dci}\cdot\mathbf{X}_{dc}\,. \tag{9}$$

Similarly, $\mathbf{V}_{dc}$ also can be extracted from $\mathbf{X}_{1\text{–}3}$ and $\mathbf{X}_{owf}$ as $\mathbf{V}_{dc} = \mathbf{F}_{dcv}\cdot[\mathbf{X}^T_{1\text{–}3}\ \mathbf{X}^T_{owf}]^T$. Then, $\mathbf{Z}_{dc}$ can be represented using $\mathbf{X}$ as:

$$\mathbf{Z}_{dc} = \begin{bmatrix} & \mathbf{F}_{dci} \\ \mathbf{F}_{dcv} & \end{bmatrix}\cdot\mathbf{X} = \mathbf{F}_{dc}\cdot\mathbf{X}\,. \tag{10}$$

Given (10), (1) can be expressed in a standard form as:

$$\begin{aligned}\dot{\mathbf{X}} &= (\mathbf{A_C} + \mathbf{E_C}\cdot\mathbf{F}_{dc})\cdot\mathbf{X} + \mathbf{B_r}\cdot\mathbf{r} + \mathbf{B_w}\cdot\mathbf{w}_d \\ &= \mathbf{A}\cdot\mathbf{X} + \mathbf{B_r}\cdot\mathbf{r} + \mathbf{B_w}\cdot\mathbf{w}_d\,.\end{aligned} \tag{11}$$

When each grid reaches a steady state after disturbance, $f_k$ and $V_{dck}$ should be restored to the nominal values. For the restoration, the MTDC-linked grid models (2) and (11) are augmented by combining the integrals of $\Delta f_k$ and $\Delta V_{dck}$ for all $k$ with $\mathbf{X}$ and $\mathbf{Y}$ as:

$$\dot{\mathbf{X}}_\mathbf{E} = \mathbf{A_E}\cdot\mathbf{X_E} + \mathbf{B_{rE}}\cdot\mathbf{r} + \mathbf{B_{wE}}\cdot\mathbf{w}_d\,, \tag{12}$$

$$\mathbf{Y_E} = \mathbf{C_E}\cdot\mathbf{X_E}\,, \tag{13}$$

with coefficient matrices augmented as:

$$\mathbf{A_E} = \begin{bmatrix} \mathbf{A} & \mathbf{O} \\ \mathbf{A_s} & \mathbf{O} \end{bmatrix}, \mathbf{B_{rE}} = \begin{bmatrix} \mathbf{B_r} \\ \mathbf{O} \end{bmatrix}, \mathbf{B_{wE}} = \begin{bmatrix} \mathbf{B_w} \\ \mathbf{O} \end{bmatrix}, \mathbf{C_E} = \begin{bmatrix} \mathbf{C} & \mathbf{O} \\ \mathbf{O} & \mathbf{I} \end{bmatrix}, \tag{14}$$

where $\mathbf{X_E} = [\mathbf{X}^T,\ \int \Delta f_k dt,\ \int \Delta V_{dck} dt]^T,\quad \forall\ k\,,$ (15)

$$\mathbf{Y_E} = [\mathbf{Y}^T,\ \int \Delta f_k dt,\ \int \Delta V_{dck} dt]^T,\quad \forall\ k\,.$$

In (14), $\mathbf{A_s}$ is a sparse matrix with elements of ones that are assigned to only the state variables $\Delta f_k$ and $\Delta V_{dck}$ for all $k$ in $\mathbf{X_E}$.

### B. *Decentralized Control Inputs and Outputs*

The proposed decentralized controllers are developed in the form of output feedback controllers, as shown in Fig. 2, so that the optimal FR in MTDC-linked grids can readily be achieved in practice using the measurement of $\mathbf{Y_E}$, rather than the estimation of $\mathbf{X_E}$. In (13), $\mathbf{Y_E}$ is decentralized to $\mathbf{Y}_{\mathbf{T}k}$ for the design of the proposed controller in grid $k$, discussed in Section III, as:

$$\begin{aligned}\mathbf{Y}_{\mathbf{T}k} &= [\mathbf{Y}_{\mathbf{E}k}^T\ \ \mathbf{Y}_{\mathbf{P}j}^T]^T = \mathbf{C}_{\mathbf{T}k}\cdot\mathbf{X_E} \\ &= [\mathbf{C}_{\mathbf{E}k}\ \ \mathbf{C}_{\mathbf{I}k}\ \ \mathbf{C}_{\mathbf{P}j}]^T\cdot\mathbf{X_E}\,,\quad \forall\ k, \text{ and } j \neq k\,.\end{aligned} \tag{16}$$

In (16), $\mathbf{Y}_{\mathbf{E}k}$ and $\mathbf{Y}_{\mathbf{P}j}$ are the measurements delivered from grids $k$ and $j$, respectively, as shown in Fig. 2. For grid $k$, $\mathbf{C}_{\mathbf{E}k}$ and $\mathbf{C}_{\mathbf{I}k}$ are obtained by decentralizing the upper and lower parts, respectively, of $\mathbf{C_E}$ in (14): i.e., $[\mathbf{C}\ \mathbf{O}] = [\mathbf{C}_{\mathbf{E}1}, \cdots, \mathbf{C}_{\mathbf{E}k}, \cdots, \mathbf{C}_{\mathbf{E}m}]^T$ and $[\mathbf{O}\ \ \mathbf{I}] = [\mathbf{C}_{\mathbf{I}1}, \cdots, \mathbf{C}_{\mathbf{I}k}, \cdots, \mathbf{C}_{\mathbf{I}m}]^T$. Moreover, in (16), $\mathbf{Y}_{\mathbf{P}j}$ is a sub-vector of $\mathbf{Y}_{\mathbf{E}j}$ and $\mathbf{C}_{\mathbf{P}j}$ is the corresponding sub-matrix of $\mathbf{C}_{\mathbf{E}j}$, because the design and activation of controller $k$ do not require

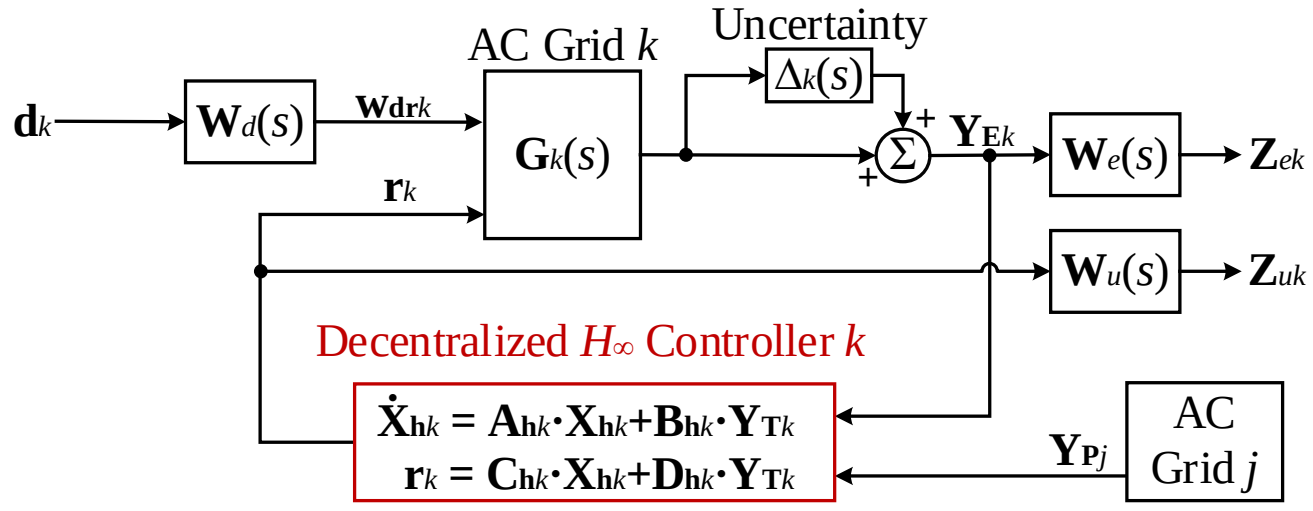

Fig. 2. A schematic diagram of the decentralized $H_\infty$ controller for ac grid $k$.

all of the measurement data in $\mathbf{Y}_{\mathbf{E}j}$, as discussed in Section III-C.

Similarly, the controller outputs or, equivalently, the reference signals $\mathbf{r}$ in (11) can be decentralized to $\mathbf{r}_k$: i.e., $\mathbf{B}_{\mathbf{rE}}\cdot\mathbf{r} = \Sigma_k \mathbf{B}_{\mathbf{rE}k}\cdot\mathbf{r}_k$. The reference signals $\mathbf{r}_j$ generated by controllers $j \neq k$ are regarded as disturbances for controller $k$. This implies that (12) can be represented from the perspective of controller $k$, as:

$$\begin{aligned}\dot{\mathbf{X}}_{\mathbf{E}} &= \mathbf{A}_{\mathbf{E}}\cdot\mathbf{X}_{\mathbf{E}} + \mathbf{B}_{\mathbf{rE}k}\cdot\mathbf{r}_k + [\mathbf{B}_{\mathbf{wE}}\ \mathbf{B}_{\mathbf{rE}j}]\cdot[\mathbf{w}_d\ \mathbf{r}_j]^T \\ &= \mathbf{A}_{\mathbf{E}}\cdot\mathbf{X}_{\mathbf{E}} + \mathbf{B}_{\mathbf{rE}k}\cdot\mathbf{r}_k + \mathbf{B}_{\mathbf{dr}k}\cdot\mathbf{w}_{\mathbf{dr}k},\ \forall\, k \text{ and } j \neq k.\end{aligned} \tag{17}$$

## III. Optimal Robust Frequency Regulation

### A. Design of an Optimal Decentralized $H_\infty$ Controller

Fig. 2 shows a closed-loop model of MTDC-linked grid $k$ with decentralized controller $k$. A general time-domain model of the decentralized controller $k$ can be established as:

$$\begin{aligned}\dot{\mathbf{X}}_{\mathbf{h}k} &= \mathbf{A}_{\mathbf{h}k}\cdot\mathbf{X}_{\mathbf{h}k} + \mathbf{B}_{\mathbf{h}k}\cdot\mathbf{Y}_{\mathbf{T}k} \\ \mathbf{r}_k &= \mathbf{C}_{\mathbf{h}k}\cdot\mathbf{X}_{\mathbf{h}k} + \mathbf{D}_{\mathbf{h}k}\cdot\mathbf{Y}_{\mathbf{T}k}\end{aligned}, \tag{18}$$

and, consequently, its frequency-domain model becomes:

$$\mathbf{r}_k = \underbrace{[\mathbf{C}_{\mathbf{h}k}\cdot(s\mathbf{I}-\mathbf{A}_{\mathbf{h}k})^{-1}\cdot\mathbf{B}_{\mathbf{h}k} + \mathbf{D}_{\mathbf{h}k}]}_{\mathbf{K}_k(s)}\cdot\mathbf{Y}_{\mathbf{T}k}. \tag{19}$$

Specifically, for the optimal robust FR in grid $k$, the controller $k$ minimizes the target performance outputs $\mathbf{Z}_{ek}$ and inputs $\mathbf{Z}_{uk}$ against disturbance $\mathbf{d}_k$. In the proposed strategy, $\mathbf{Z}_{ek}$ includes not only the weighted values of $\Delta f_k$ and $\int\Delta f_k\, dt$ but also the weighted values of $\Delta V_{dck}$ and $\int\Delta V_{dck}\, dt$ to prevent excessive variations in dc terminal voltages during the optimal FR. Similarly, $\mathbf{Z}_{uk}$ contains the weighted values of $\Delta P_{gk}^{ref}$, $\Delta P_k^{ref}$, and $\Delta V_{dck}^{ref}$ when grid $k$ is linked via a VSC. For the case of an LCC, it consists of $\Delta P_{gk}^{ref}$ and $\Delta P_k^{ref}$. Furthermore, $\mathbf{W}_e$, $\mathbf{W}_u$, and $\mathbf{W}_d$ are the weighting functions for $\mathbf{Z}_{ek}$, $\mathbf{Z}_{uk}$, and $\mathbf{d}_k$, respectively, as:

$$\mathbf{Z}_k = \begin{bmatrix}\mathbf{Z}_{ek}\\ \mathbf{Z}_{uk}\end{bmatrix} = \begin{bmatrix}\mathbf{W}_e\cdot\bar{\mathbf{C}}_{\mathbf{E}k} & \mathbf{O}\\ \mathbf{O} & \mathbf{W}_u\cdot\mathbf{I}\end{bmatrix}\begin{bmatrix}\mathbf{X}_{\mathbf{E}}\\ \mathbf{r}_k\end{bmatrix} = \begin{bmatrix}\mathbf{C}_1 & \mathbf{D}_{12}\end{bmatrix}\begin{bmatrix}\mathbf{X}_{\mathbf{E}}\\ \mathbf{r}_k\end{bmatrix}, \tag{20}$$

$$\text{and}\quad \mathbf{w}_{\mathbf{dr}k} = \mathbf{W}_d\cdot\mathbf{d}_k,$$

where $\bar{\mathbf{C}}_{\mathbf{E}k}$ is a sparse matrix with elements of ones to extract $\Delta f_k$, $\Delta V_{dck}$, $\int\Delta f_k\, dt$, and $\int\Delta V_{dck}\, dt$ from $\mathbf{X}_{\mathbf{E}}$. Note that in (20), $\mathbf{C}_1$ and $\mathbf{D}_{12}$ are additionally defined for brief notation. Moreover, in Fig. 2, $\Delta_k(s)$ reflects the uncertainty in the model parameters of grid $k$ [20], [21]: i.e., $G_k(s)\cdot(1+\Delta_k(s))$, rather than $G_k(s)$. In time domain, $\Delta_k(s)$ affects the accuracy of $\mathbf{A}_{\mathbf{E}}$, $\mathbf{B}_{\mathbf{rE}k}$, $\mathbf{B}_{\mathbf{dr}k}$, and $\mathbf{C}_{\mathbf{T}k}$ [see (16) and (17)]. This implies that the proposed decentralized controllers are developed considering uncertainty in the model parameter estimates of SGs, converters, and ac and dc transmission lines.

Given the weighting functions and parameter uncertainty, the model of the MTDC-linked grid with the decentralized control inputs and outputs [i.e., (16) and (17)] can be completed as:

$$\dot{\mathbf{X}}_{\mathbf{E}} = \mathbf{A}_1\cdot\mathbf{X}_{\mathbf{E}} + \mathbf{B}_1\cdot\mathbf{d}_k + \mathbf{B}_2\cdot\mathbf{r}_k, \tag{21}$$

$$\mathbf{Z}_k = \mathbf{C}_1\cdot\mathbf{X}_{\mathbf{E}} + \mathbf{D}_{12}\cdot\mathbf{r}_k, \tag{22}$$

$$\mathbf{Y}_{\mathbf{T}k} = \mathbf{C}_2\cdot\mathbf{X}_{\mathbf{E}}, \tag{23}$$

where the notations are simplified as $\mathbf{A}_1 = \mathbf{A}_{\mathbf{E}}$, $\mathbf{B}_1 = \mathbf{B}_{\mathbf{dr}k}\cdot\mathbf{W}_d$, $\mathbf{B}_2 = \mathbf{B}_{\mathbf{rE}k}$, and $\mathbf{C}_2 = \mathbf{C}_{\mathbf{T}k}$. In frequency domain, (21) is expressed as:

$$\mathbf{X}_{\mathbf{E}} = (s\mathbf{I}-\mathbf{A}_1)^{-1}\cdot[\mathbf{B}_1\ \mathbf{B}_2]\cdot[\mathbf{d}_k\ \mathbf{r}_k]^T. \tag{24}$$

Using (24), (22) and (23) are then equivalently represented in the frequency domain as:

$$\mathbf{Z}_k = \{\mathbf{C}_1\cdot(s\mathbf{I}-\mathbf{A}_1)^{-1}\cdot[\mathbf{B}_1\ \mathbf{B}_2] + [\mathbf{O}\ \mathbf{D}_{12}]\}\cdot[\mathbf{d}_k\ \mathbf{r}_k]^T, \tag{25}$$

$$\text{and}\quad \mathbf{Y}_{\mathbf{T}k} = \mathbf{C}_2\cdot(s\mathbf{I}-\mathbf{A}_1)^{-1}\cdot[\mathbf{B}_1\ \mathbf{B}_2]\cdot[\mathbf{d}_k\ \mathbf{r}_k]^T. \tag{26}$$

In other words, for the optimal robust FR, decentralized controller $k$ [see (18) and (19)] should be designed to generate $\mathbf{r}_k$ that minimizes $\mathbf{Z}_k$ for all unknown $\mathbf{d}_k$ in (25), given the measurements $\mathbf{Y}_{\mathbf{T}k}$ in (26).

In the proposed FR strategy, the optimal controller gain $\mathbf{K}_k(s)$ is determined to minimize the $H_\infty$-norm of the transfer function $\mathbf{T}_{\mathbf{Z}k\mathbf{d}k}$ from $\mathbf{d}_k$ to $\mathbf{Z}_k$ by solving the robust optimization problem:

$$\min\left\|\mathbf{T}_{\mathbf{Z}k\mathbf{d}k}(\mathbf{K}_k(s))\right\|_\infty < \gamma, \tag{27}$$

$$\text{where}\quad \left\|\mathbf{T}_{\mathbf{Z}k\mathbf{d}k}(\mathbf{K}_k(s))\right\|_\infty = \sup_\omega \bar{\sigma}(\mathbf{T}_{\mathbf{Z}k\mathbf{d}k}(j\omega)), \tag{28}$$

$$\begin{aligned}\mathbf{T}_{\mathbf{Z}k\mathbf{d}k}(s) = \mathbf{G}_{11}(s) + \mathbf{G}_{12}(s)\mathbf{K}_k(s)\cdot \\ (\mathbf{I}-\mathbf{G}_{22}(s)\mathbf{K}_k(s))^{-1}\mathbf{G}_{21}(s)\end{aligned}, \tag{29a}$$

$$\mathbf{G}_{11}(s) = \mathbf{C}_1\cdot(s\mathbf{I}-\mathbf{A}_1)^{-1}\cdot\mathbf{B}_1, \tag{29b}$$

$$\mathbf{G}_{12}(s) = \mathbf{C}_1\cdot(s\mathbf{I}-\mathbf{A}_1)^{-1}\cdot\mathbf{B}_2 + \mathbf{D}_{12}, \tag{29c}$$

$$\mathbf{G}_{21}(s) = \mathbf{C}_2\cdot(s\mathbf{I}-\mathbf{A}_1)^{-1}\cdot\mathbf{B}_1, \tag{29d}$$

$$\text{and}\quad \mathbf{G}_{22}(s) = \mathbf{C}_2\cdot(s\mathbf{I}-\mathbf{A}_1)^{-1}\cdot\mathbf{B}_2. \tag{29e}$$

In (27) and (28), $\mathbf{T}_{\mathbf{Z}k\mathbf{d}k}$ can be obtained as (29) by substituting (19) into (24)–(26). Moreover, $\bar{\sigma}$ represents a maximum singular value of $\mathbf{T}_{\mathbf{Z}k\mathbf{d}k}$ and $\gamma$ is the performance level. In this study, a $\gamma$-iteration method is adopted to solve the optimization problem (27)–(29), wherein $\gamma$ is repeatedly updated to a smaller value until convergence [22]. For each value of $\gamma$, there exists a stable $H_\infty$ controller that satisfies (27) if and only if the three conditions hold [23] as:

$$\mathbf{X}_\infty = -\mathbf{A}_1\mathbf{A}_1^T + \mathbf{C}_1^T\mathbf{C}_1(\gamma^{-2}\mathbf{B}_1\mathbf{B}_1^T - \mathbf{B}_2\mathbf{B}_2^T) \geq 0, \tag{30a}$$

$$\mathbf{Y}_\infty = -\mathbf{A}_1^T\mathbf{A}_1 + \mathbf{B}_1\mathbf{B}_1^T(\gamma^{-2}\mathbf{C}_1^T\mathbf{C}_1 - \mathbf{C}_2^T\mathbf{C}_2) \geq 0, \tag{30b}$$

$$\text{and}\quad \rho(\mathbf{X}_\infty\mathbf{Y}_\infty) < \gamma^2. \tag{30c}$$

In (30c), $\rho(\mathbf{X}_\infty\mathbf{Y}_\infty)$ is the largest eigenvalue of $\mathbf{X}_\infty\mathbf{Y}_\infty$. When $\gamma$ converges, the optimal controller gain $\mathbf{K}_k(s)$ is then determined using (30), as:

$$\mathbf{K}_k(s) = \begin{bmatrix}\mathbf{A}_\infty & -\mathbf{Z}_\infty\mathbf{L}_\infty\\ \mathbf{F}_\infty & 0\end{bmatrix}, \tag{31a}$$

$$\text{where}\quad \mathbf{A}_\infty = \mathbf{A}_1 + \gamma^{-2}\mathbf{B}_1\mathbf{B}_1^T\mathbf{X}_\infty + \mathbf{B}_2\mathbf{F}_\infty + \mathbf{Z}_\infty\mathbf{L}_\infty\mathbf{C}_2, \tag{31b}$$

$$\mathbf{F}_\infty = -\mathbf{B}_2^T\mathbf{X}_\infty,\ \mathbf{L}_\infty = -\mathbf{Y}_\infty\mathbf{C}_2^T,\ \text{and}\ \mathbf{Z}_\infty = (\mathbf{I}-\gamma^{-2}\mathbf{Y}_\infty\mathbf{X}_\infty)^{-1}. \tag{31c}$$

From (19) and (31), the time-domain control gains $\mathbf{A}_{\mathbf{h}k}$, $\mathbf{B}_{\mathbf{h}k}$, $\mathbf{C}_{\mathbf{h}k}$, and $\mathbf{D}_{\mathbf{h}k}$ in (18) also can be obtained, which is omitted here.

### B. Model Order Reduction of Decentralized $H_\infty$ Controllers

In general, a dynamic model of MTDC-linked grids [i.e., (12)–(15)] is implemented in high dimensions including several uncontrollable and unobservable variables particularly when the numbers of SGs, VSCs, LCCs, and WTGs are large. For the case of Grids 1–3, discussed in Section II-A, each of three decentralized $H_\infty$ controllers is designed with a model order $r$ of 87; in other words, the size of $\mathbf{X}_{hk}$ in (18) is 87 for each $k$. The full-order $H_\infty$ controllers are difficult to implement in practice due to high computational burden and high sensitivity to measurement noises. In this paper, a balanced truncation algorithm [24] is applied to reduce the model orders of the $H_\infty$ controllers while still preserving their dominant response characteristics, thus facilitating practical applications of the proposed strategy.

Fig. 3(a) shows the Hankel singular values (HSVs) of the full-order and reduced-order models of the three decentralized $H_\infty$ controllers for the MTDC-linked grids (i.e., Grids 1–3). The HSVs were obtained for the step response to the load demand variation by 0.1 pu in each grid, given the test conditions discussed in Sections IV-A and IV-C. Fig. 3(b) shows the corresponding cumulative energy curves. The results of the HSV analysis indicate that the energy magnitudes of the three decentralized $H_\infty$ controllers are still higher than 99.9% of the total energy when their model orders are reduced to $r$ = 26, 29, and 13, respectively. This implies that the reduced-order models of the $H_\infty$ controllers can successfully reflect the dynamic operating characteristics of the original, 87$^{th}$-order models. In other words, the reduction of the model orders to 26, 29, and 13, respectively, only marginally affects the FR performance, while effectively mitigating the computational burden. Note that $r$ can be set to smaller values, considering the trade-off between control performance and implementation complexity. For example, when the model order of the $H_\infty$ controller for Grid 1 is reduced to $r$ = 10 and 2, rather than $r$ = 26, the HSV becomes higher than 0.014 and 0.414, respectively. The cumulative energy then decreases to 98.8% and 52.7%, respectively.

## *C. Sensitivity and Eigenvalue Analyses*

The closed-loop system, shown in Fig. 2, was analyzed with respect to the gain sensitivity and stability. Note that the analysis was conducted using reduced-order models of the three decentralized $H_\infty$ controllers, discussed in Section III-B. Specifically, Fig. 4 shows the optimal gains of the $H_\infty$ controllers for the GSVSC- and LCC-interfaced grids (i.e., $k$ = 1, 2, and 3) particularly with regard to the measurements from the corresponding grid and the interconnected grids: i.e., $\mathbf{B}_{hk}$ and $\mathbf{D}_{hk}$ for $\mathbf{Y}_{Tk} = [\mathbf{Y}^T_{Ek}\ \mathbf{Y}^T_{Pj}]^T$ in Fig. 2 and (18). It represents the extent to which the measurements $\mathbf{Y}_{Tk}$ affect the controller states $\mathbf{X}_{hk}$ and the reference signals $\mathbf{r}_k$ under disturbed grid conditions. It can be seen that in each grid $k$, the elements in $\mathbf{B}_{hk}$ and $\mathbf{D}_{hk}$ have values close to zero for all the measurements from grid $j \neq k$, apart from the measurements of $\Delta f_j$ and $\Delta V_{dcj}$. In other words, $\mathbf{X}_{hk}$ and $\mathbf{r}_k$ in each grid $k$ are mainly affected by the measurements of $\Delta f_j$ and $\Delta V_{dcj}$ for grid $j \neq k$. This demonstrates that the decentralized $H_\infty$ controller for grid $k$ can be effectively designed only using the measurements of $\Delta f_j$ and $\Delta V_{dcj}$ for grid $j \neq k$ [see (A16) for $\mathbf{Y}_j$], thus mitigating the requirement for inter-grid measurement and communication systems and facilitating wide applications of the proposed strategy in practice.

In addition, Fig. 5 shows the eigenvalues of the closed-loop system with changes in system parameters, model uncertainty level, and communication time delay. In Fig. 5(a), as the phase inductance $L_f$ at the PCC of each MTDC converter increases from 0.03 mH to 1.50 mH, the complex-conjugate eigenvalues move closer to the imaginary axis, implying increased overshoot and settling time. However, all eigenvalues are still placed in the left-hand half plane (LHP), verifying that the proposed decentralized $H_\infty$ controllers ensure grid frequency stability for a large variation in $L_f$. Note that the eigenvalues are more affected by $L_f$, than the other system parameters. Analogously, Fig. 5(b) shows the loci of the eigenvalues when the uncertainty level $\Delta(s)$ of the system model parameters increases from 0% to 50%. The dominant, complex-conjugate eigenvalues are still located in the LHP for a large value of $\Delta(s)$ (i.e., up to $\Delta(s)$ = 47%). This confirms that the proposed $H_\infty$ controllers can secure the stable operation of MTDC-linked grids even when there is relatively large uncertainty in the system parameter estimates. Moreover, Fig 5(c) shows the eigenvalue analysis for an increase in $T_d$ from 0 s to 0.6 s; for brevity, the delay is assumed to be constant and identical for all decentralized $H_\infty$ controllers. Given $T_d$, (18) changes to:

$$\begin{aligned} \dot{\mathbf{X}}_{\mathbf{h}k}(t) &= \mathbf{A}_{\mathbf{h}k}\cdot\mathbf{X}_{\mathbf{h}k}(t)+\mathbf{B}_{\mathbf{h}k}\cdot\mathbf{Y}_{\mathbf{T}k}(t-T_d) \\ \mathbf{r}_k(t) &= \mathbf{C}_{\mathbf{h}k}\cdot\mathbf{X}_{\mathbf{h}k}(t)+\mathbf{D}_{\mathbf{h}k}\cdot\mathbf{Y}_{\mathbf{T}k}(t-T_d) \end{aligned}. \quad (32)$$

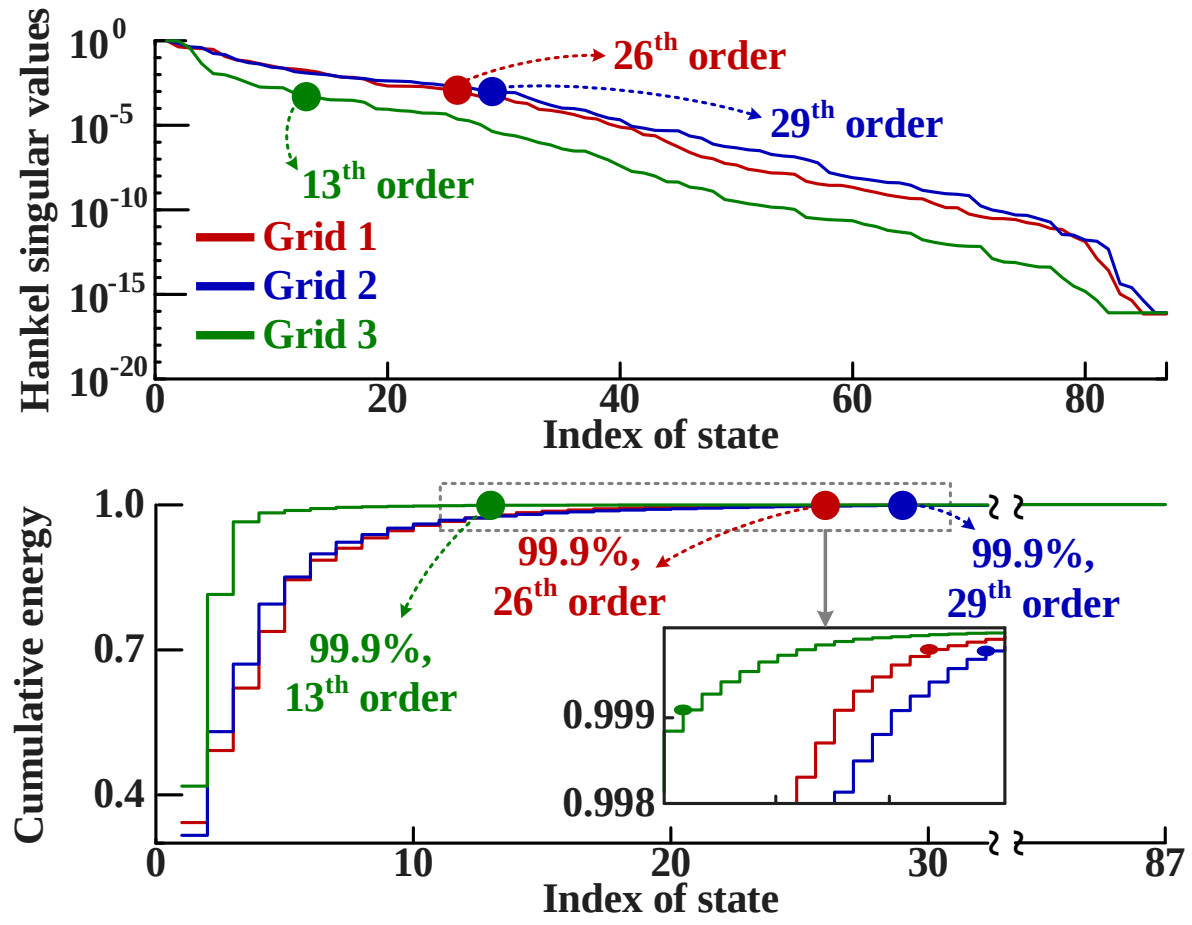

Fig. 3. (a) Hankel singular values and (b) cumulative energy curves for the optimal decentralized $H_\infty$ controllers.

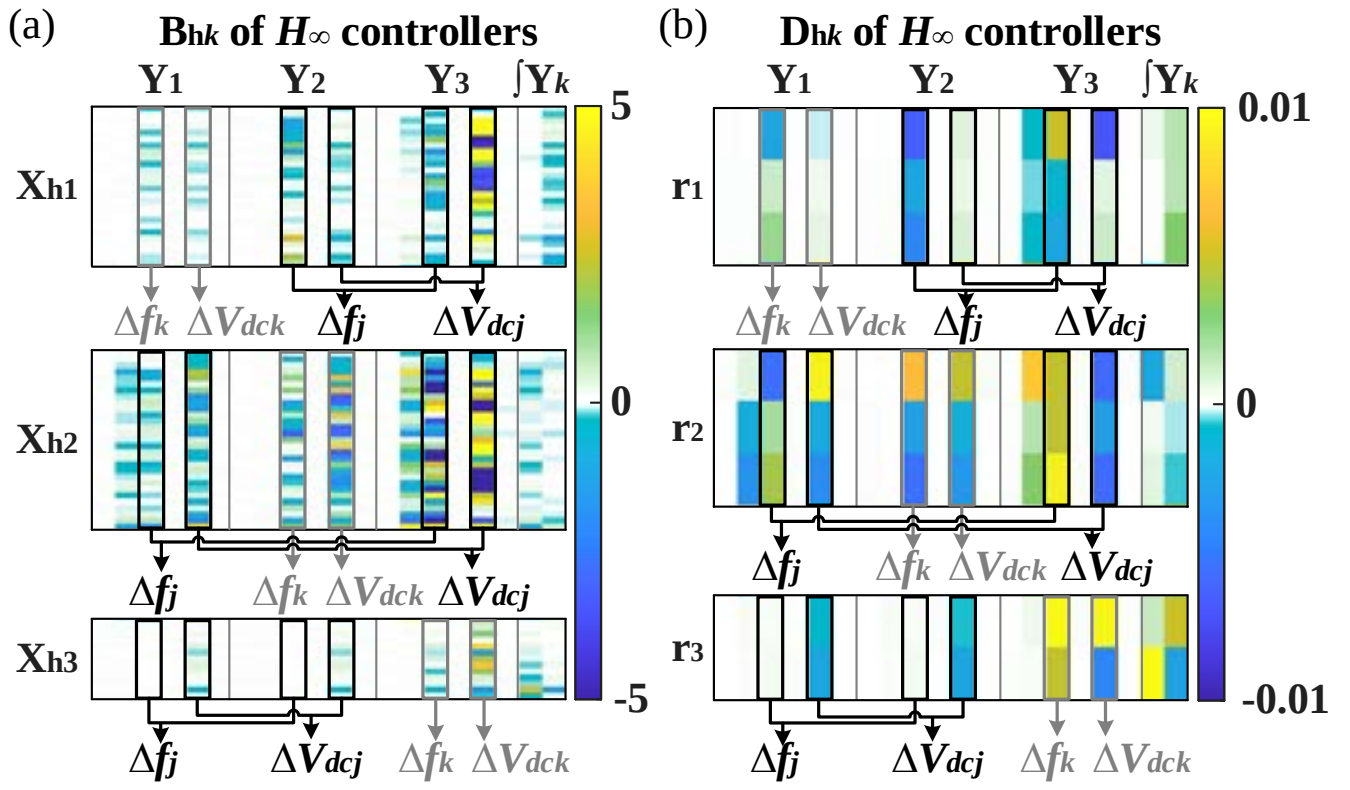

Fig. 4. Gains of the optimal decentralized $H_\infty$ controllers for grids $k$ =1, 2, and 3: (a) $\mathbf{B}_{hk}$ and (b) $\mathbf{D}_{hk}$ for all $k$ and $j \neq k$.

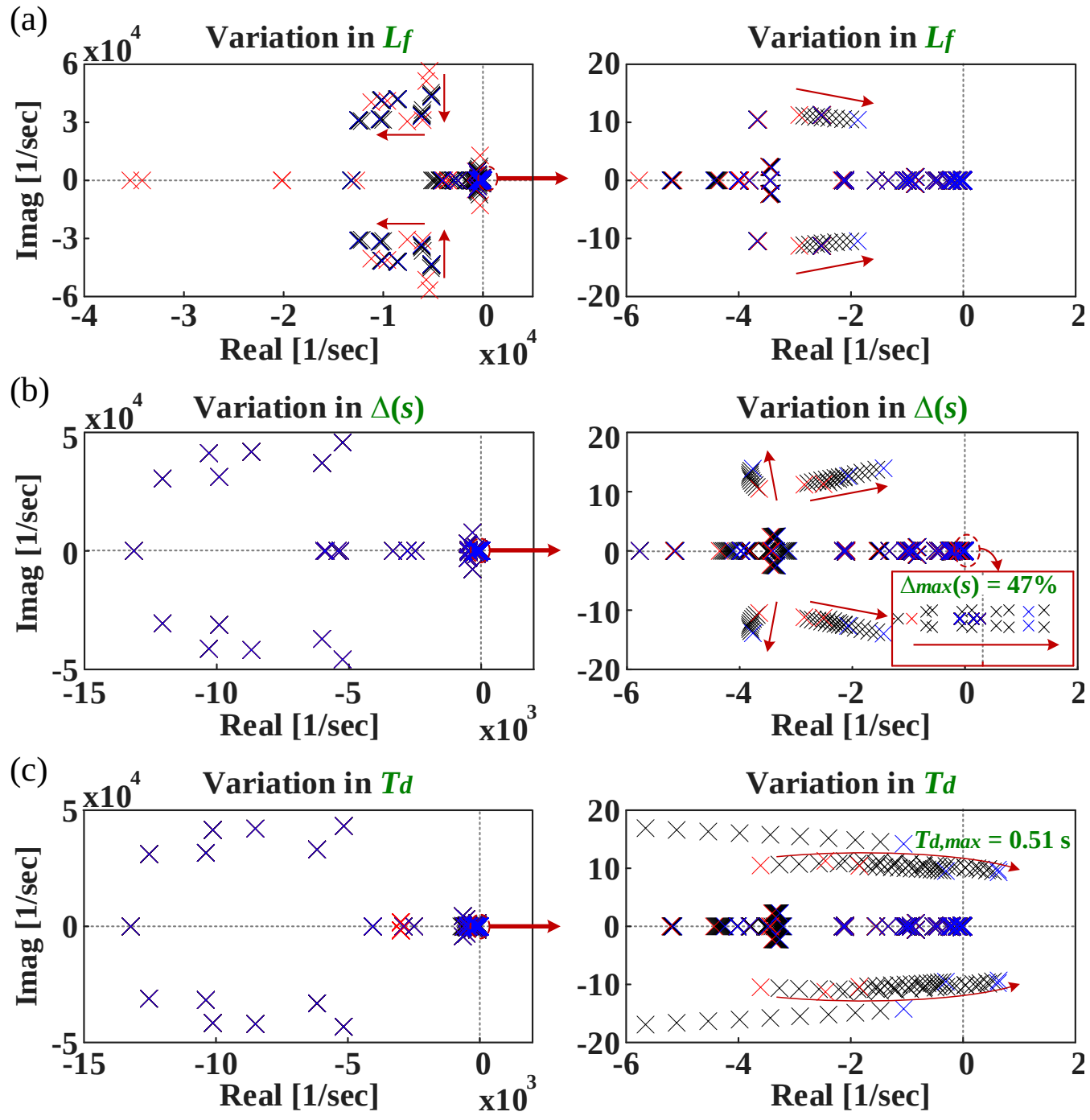

Fig. 5. Closed-loop system eigenvalues for variations in (a) $L_f$ ranging from 0.03 mH to 1.50 mH, (b) $\Delta(s)$ from 0% to 50%, and (c) $T_d$ from 0 s to 0.6 s.

Therefore, in the frequency domain, (19) is replaced by:

$$\mathbf{r}_k(s)=\mathbf{K}_k(s)\cdot e^{-T_d s}\cdot\mathbf{Y}_{\mathbf{T}k}(s)\,, \quad (33)$$

where $e^{-T_d s}$ can be approximated as:

$$e^{-T_d s}\approx\frac{T_d^2 s^2-6T_d s+12}{T_d^2 s^2+6T_d s+12}. \quad (34)$$

In Fig. 5(c), with an increase in $T_d$, the complex-conjugate eigenvalues move toward the imaginary axis. When $T_d$ increases to approximately 0.51 s, the eigenvalues cross the axis and are placed on the right-hand half plane (RHP). It was reported in [25] that the time delay in common fiber optic communications for application to MTDC systems ranges between 1 ms and 10 ms. Thus, the proposed $H_\infty$ controllers can successfully guarantee the stability and robustness of the closed-loop system in practice.

## IV. Case Studies and Results

### *A. Test System and Simulation Conditions*

For comparative case studies, a dynamic model of hybrid MTDC-linked grids, discussed in Section II-A, was implemented in MATLAB/SIMULINK (or, simply, SIMULINK). Each ac grid consisted mainly of SGs, transmission lines, regional loads, and a GSVSC or LCC station for the interface with the MTDC system. In SIMULINK, the hybrid MTDC system was modeled using average circuit models of the GSVSC, WFVSC, and LCC with inner-loop and outer-loop feedback controllers in a *dq* frame. Moreover, a Π-section model of the dc link was included to interconnect the MTDC converters, forming the dc transmission network. An OWF model was also implemented using dynamic models of wind turbines and PMSGs and then connected to the dc network via a WFVSC. Table I shows the model parameters of the SGs, MTDC converters, ac and dc networks, and PMSG-type OWF, as well as the corresponding controller gains at the local level [26]–[28]. Fig. 6 shows the load demand variations $\Delta P_{Lk}$ for all grids $k$, reflecting the scaled-down RegD signals [29] over a period of 200 s. It also shows the intermittent wind speed for the OWF over 200 s.

Given the test bed parameters and operating conditions, the proposed decentralized $H_\infty$ controllers were developed with no loop shaping (i.e., unity weighting factors) [20] and analyzed in comparison with conventional decentralized controllers [19], [30]; the effects of different weighting functions such as low-pass filters (LPFs) and band-pass filters (BPFs) were also analyzed. Table II shows the main features of the proposed (Case 1) and conventional strategies (Case 2 and 3) for decentralized FR of the MTDC-linked grids. In Case 2, a decentralized PI controller was adopted to regulate the regional grid frequency deviations without the coordination between SGs and MTDC converters. In Case 3, a centralized optimization problem was formulated to determine the decentralized $H_\infty$ controller gains, unlike Case 1 wherein the decentralized optimization problem for each grid was directly formulated. Specifically, the solution to the centralized problem led to the centralized $H_\infty$ control gains and, consequently, the decentralized controllers were implemented by setting the off-diagonal elements of the centralized gains to zero, excluding the elements corresponding to $\Delta f_k$ and $\Delta V_{dck}$ in each grid $k$. This implies that the decentralized FR of the MTDC-linked grids was not truly optimized in Case 3.

TABLE I. SYSTEM PARAMETERS FOR CASE STUDIES

| Device | Parameter | | Value | Unit |
|---|---|---|---|---|
| Synchronous generators | Rated voltage (*ph–ph*) | $V_{rate}$ | 26.30 | [kV] |
| | Moment of inertia | $H$ | 3.5 | [s] |
| | Stator impedance | $R_s$, $L_s$ | 0.0015, 0.15 | [pu] |
| | Transient open time constants | $T'_{do}$, $T'_{qo}$ | 2.00, 0.75 | [s] |
| | Sub-transient open time constants | $T''_{do}$, $T''_{qo}$ | 0.30, 0.055 | [s] |
| Governors and turbines | Droop gain | $R$ | 0.05 | [pu] |
| | Time constants | $T_g$, $T_{rh}$ | 0.01, 7 | [s] |
| | Power fraction | $F_{hp}$ | 0.3 | [pu] |
| Exciters | Gain and time constant | $K_c$, $T_a$ | 200, 0.02 | [pu] |
| MTDC converters | Phase resistance and inductance | $R_f$, $L_f$ | 0.003, 0.05 | [pu] |
| | GSVSC capacitance | $C_{vsc}$ | 200 | [μF] |
| | LCC smoothing inductance | $L_{sm}$ | 0.02 | [pu] |
| | LCC filter capacitance | $C_f$ | 0.1 | [pu] |
| AC and DC networks | Rated voltages | $V_{rate,ac}$<br>$V_{rate,dc}$ | 380,<br>±320 | [kV] |
| | Transformer inductance | $L_g$ | 0.01 | [pu] |
| | DC cable impedance | $R_{dc}$,<br>$L_{dc}$,<br>$C_{dc}$, | 0.0139,<br>0.159,<br>0.231 | [Ω/km]<br>[mH/km]<br>[μF/km] |
| PMSG-type OWF | Moment of inertia | $H$ | 0.685 | [s] |
| | Pole pairs | $p$ | 48 | [pairs] |
| | Stator impedance | $R_s$, $L_s$ | 0.027, 0.5131 | [pu] |
| | Magnetic flux | $\varphi_f$ | 1.1884 | [pu] |
| PI controllers | PI gains of GSVSC | $K_p$, $K_i$ | 0.5, 2 | [pu] |
| | Droop gain of GSVSC | $R_{droop}$ | 2 | [pu] |
| | PI gains of LCC | $K_p$, $K_i$ | 0.7, 5 | [pu] |
| | PI gains of WFVSC | $K_p$, $K_i$ | 0.5. 3 | [pu] |

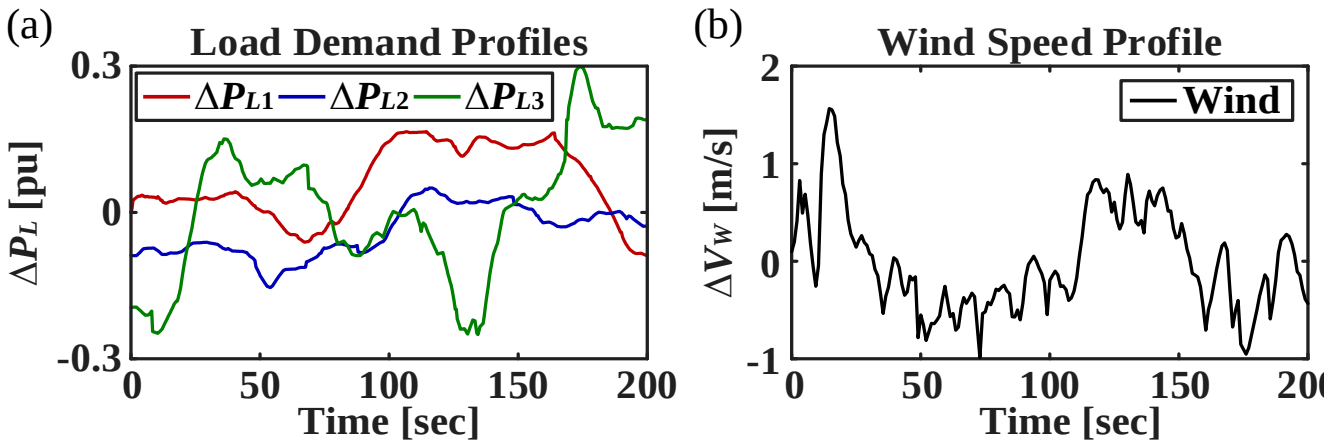

Fig. 6. Continuous variations in the (a) loads $\Delta P_{L1\text{–}3}$ and (b) wind speed $\Delta V_w$.

TABLE II. FEATURES OF THE PROPOSED AND CONVENTIONAL STRATEGIES

| Strategies | | Optimization Problem | Controller Structure | Optimal Gains | Device Coordination |
|---|---|---|---|---|---|
| Prop. | Case 1 $H_\infty$ | Decentralized | Decentralized | O | O |
| Conv. | Case 2 PI | – | Decentralized | – | X |
| | Case 3 $H_\infty$ | Centralized | Decentralized | X | O |

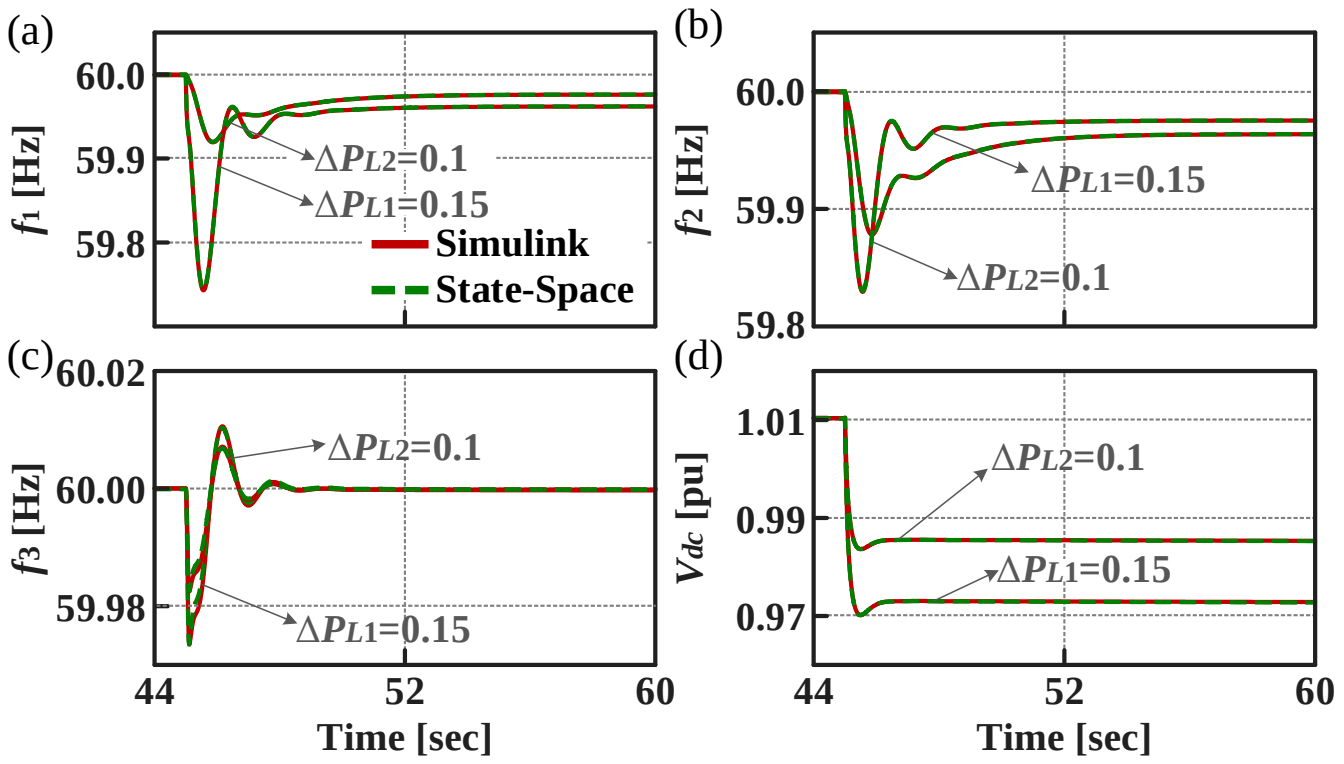

Fig. 7. Comparisons of the step responses of the small-signal model and the comprehensive SIMULINK model of the MTDC-linked grids to $\Delta P_{L1}$ = 0.1 pu and $\Delta P_{L2}$ = 0.15 pu: (a) $f_1$, (b) $f_2$, (c) $f_3$, and (d) $V_{dc}$.

## *B. Validating the Dynamic Model of MTDC-linked Grids*

Fig. 7 shows comparisons between the responses of two models of the MTDC-linked grids to a step increase in $\Delta P_{L1}$ = 0.1 pu and $\Delta P_{L2}$ = 0.15 pu at $t$ = 45 s: i.e., the small-signal model and the comprehensive SIMULINK model, discussed in Sections II and IV-A, respectively. The small-signal model was developed in a linear time-invariant form, whereas the SIMULINK model was implemented in a nonlinear time-varying form. Note that the comparisons were conducted only for MTDC-linked grids with local droop controllers of SGs and MTDC converters, rather than grids with the proposed and conventional secondary frequency controllers (i.e., Cases 1–3). In Fig. 7, the dynamic responses of the two models were very similar to each other for each profile of the grid frequencies $f_{1\text{–}3}$ and the average dc voltage $V_{dc}$ for all MTDC terminals. This indicates good consistency between the two models even when relatively large variations in load demands occur in the MTDC-linked grids under normal conditions. It hence validates the accuracy of the results obtained by applying the proposed and conventional controllers to the small-signal dynamic model, discussed in Section IV-C, D, and E. Moreover, Fig. 7(a)–(c) shows that in each grid, the frequency fell off and then reached to a steady state value with a large overshoot and settling time. This confirms the motivation of this study on the coordinated FR of MTDC-linked grids. Note that $f_3$ was maintained at 60 Hz in the steady state, because the LCC operated with a constant power reference, unlike the GSVSCs with $P$-$V_{dc}$ droop controllers.

## *C. Comparisons for Stepwise Load Variations*

Fig. 8 shows $f_{1\text{–}3}$ and $V_{dc}$ for the step responses of the MTDC-linked grids to load variations $\Delta P_{Lk}$, which increased by 0.1 pu for all $k$ at $t$ = 10 s over a 10 s period. Table III lists the corresponding numerical results. In the proposed strategy (i.e., Case 1), the frequency deviations became smaller for all grids $k$ than in the conventional strategies (i.e., Cases 2 and 3). Specifically, Case 1 led to a decrease in the sum of the maximum frequency deviations (i.e., $\Sigma_k\,|\Delta f_k|_{\max}$) by 75.7% and 8.46%, compared to Cases 2 and 3, respectively. Moreover, in Case 1, $f_{1\text{–}3}$ were restored back to the nominal values more rapidly, and had smaller overshoot responses than in Cases 2 and 3. Case 1 also resulted in smaller $\Delta V_{dc}$ by 13.0% and 42.9% than Cases 2 and 3, respectively, as shown in Fig. 8(d). In other words, the proposed decentralized $H_\infty$ controllers were more effective in reducing the frequency deviations in all the MTDC-linked grids while further mitigating the dc voltage variations. The improvement was made only using the measurements of $\Delta f_j$ and $\Delta V_{dcj}$, rather than all the measurements in grid $j \neq k$, thus lessening the requirement on inter-grid sensing and communication systems.

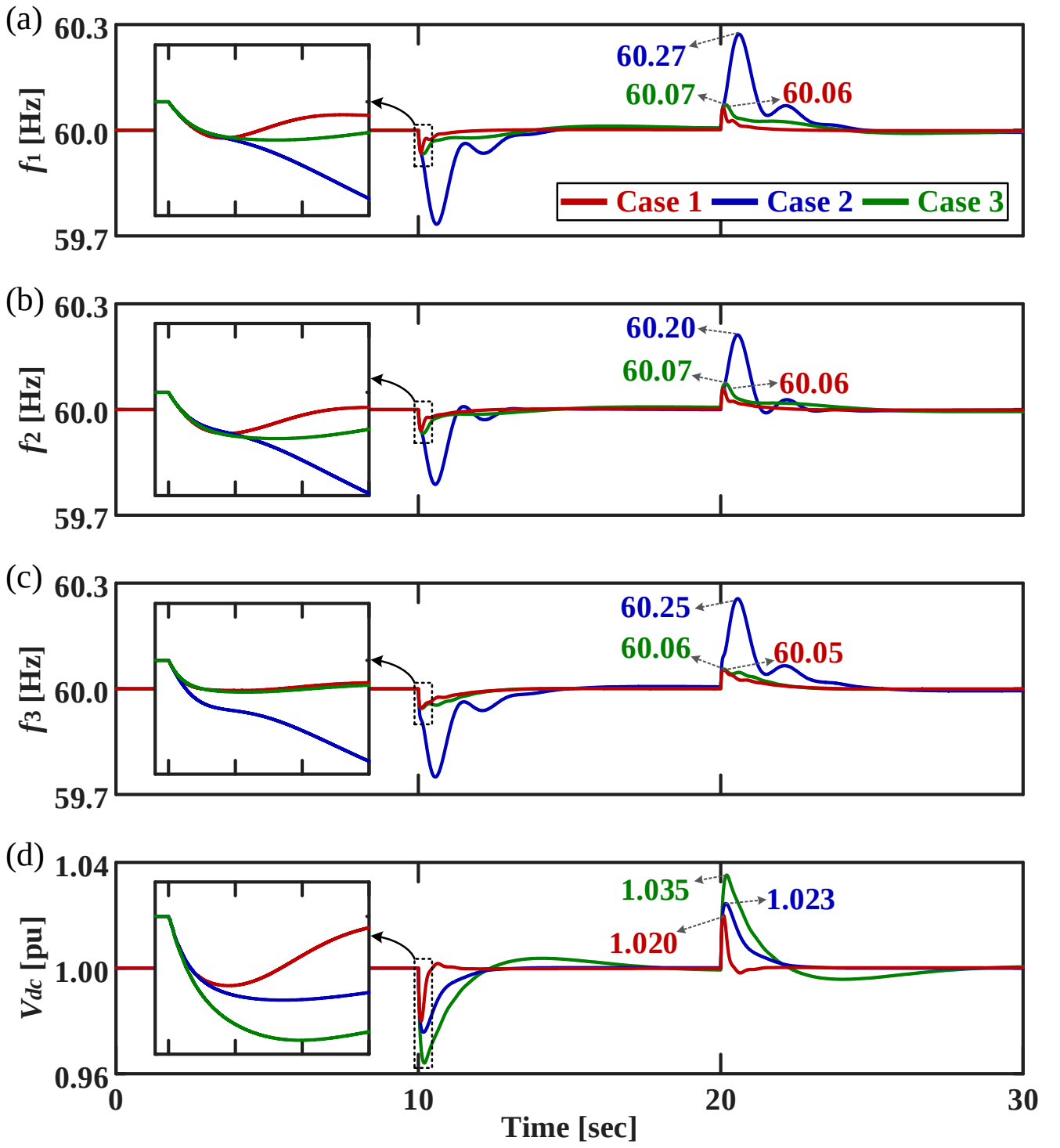

Fig. 8. Step responses of the proposed and conventional strategies: (a) $f_1$, (b) $f_2$, (c) $f_3$, and (d) $V_{dc}$.

TABLE III. COMPARISONS OF STEP RESPONSE TEST RESULTS

| Maximum Variations | Case 1 Individual | Case 1 Total | Case 2 Individual | Case 2 Total | Case 3 Individual | Case 3 Total |
|---|---|---|---|---|---|---|
| $|\Delta f_1|_{\max}$ [Hz] | 0.064 | | 0.273 | | 0.070 | |
| $|\Delta f_2|_{\max}$ [Hz] | 0.060 | 0.178 | 0.207 | 0.730 | 0.068 | 0.194 |
| $|\Delta f_3|_{\max}$ [Hz] | 0.053 | | 0.250 | | 0.056 | |

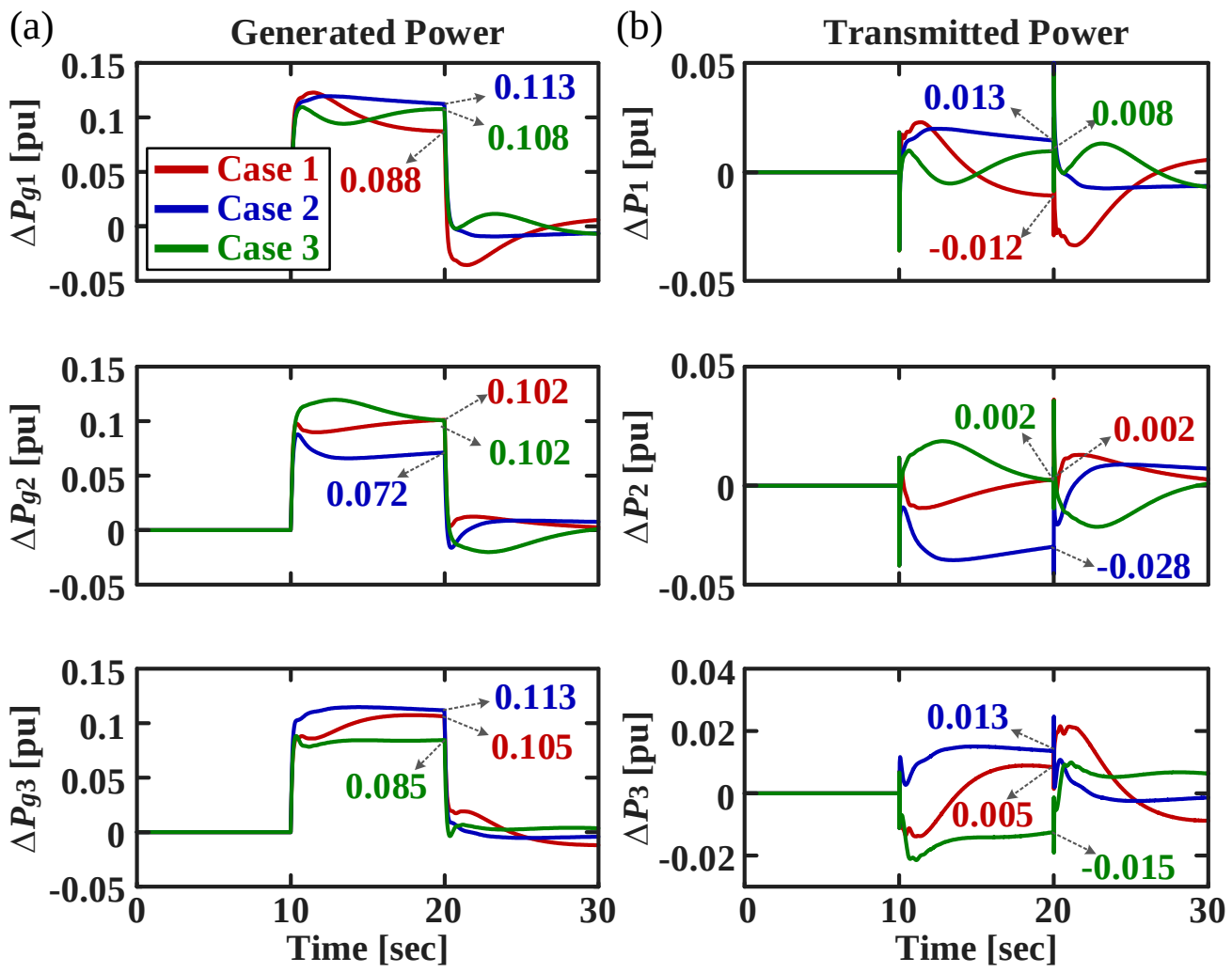

Fig. 9. Corresponding variations in (a) the generated and (b) transmitted power.

In addition, Fig. 9(a) shows the profiles of total power generation in each grid to stepwise load variations. Case 1 led to only moderate variations in generated power. This is also the case for the power transmitted by MTDC converters, as shown in Fig. 9(b). The comparison results demonstrate that the improved performance of Case 1 is attributable to the optimal, coordinated power sharing among the MTDC-linked grids in transient and steady states, rather than excessive variations in the generated and transmitted power.

### *D. Comparisons for Continuous Load and Wind Variations*

The proposed FR strategy was also tested with regard to continuous variations in the grid load demands $\Delta P_{L1\text{–}3}$ and wind speed $\Delta V_w$ over a time period of 200 s, as shown in Fig. 6. Fig. 10 shows $f_{1\text{–}3}$ and $V_{dc}$ for the responses of the MTDC-linked grids to $\Delta P_{L1\text{–}3}$ and $\Delta V_w$, and Table IV lists the numerical results as root-mean-square (rms) values. Each rms value was calculated as $\{\sum_l \Delta f_k(l)^2/L\}^{1/2}$ and $\{\sum_l \Delta V_{dc}(l)^2/L\}^{1/2}$, where $l$ is the index of the measurement samples and $L$ is the total number of samples. Case 1 led to significantly smaller variations in the frequencies and dc terminal voltages for all the grids than Cases 2 and 3. Specifically, the sum of $\Delta f_{1\text{–}3,rms}$ in Case 1 was 92.5% and 62.5% smaller than those in Case 2 and 3, respectively. Moreover, in Case 1, $\Delta V_{dc,rms}$ was reduced by 66.7% and 75.0%, compared to Cases 2 and 3. The case study results consistently verify that the proposed decentralized $H_\infty$ controllers more effectively improve the FR of the MTDC-linked grids while further suppressing the dc network voltage variations through optimal, coordinated sharing of the time-varying power outputs of the SGs and OWF.

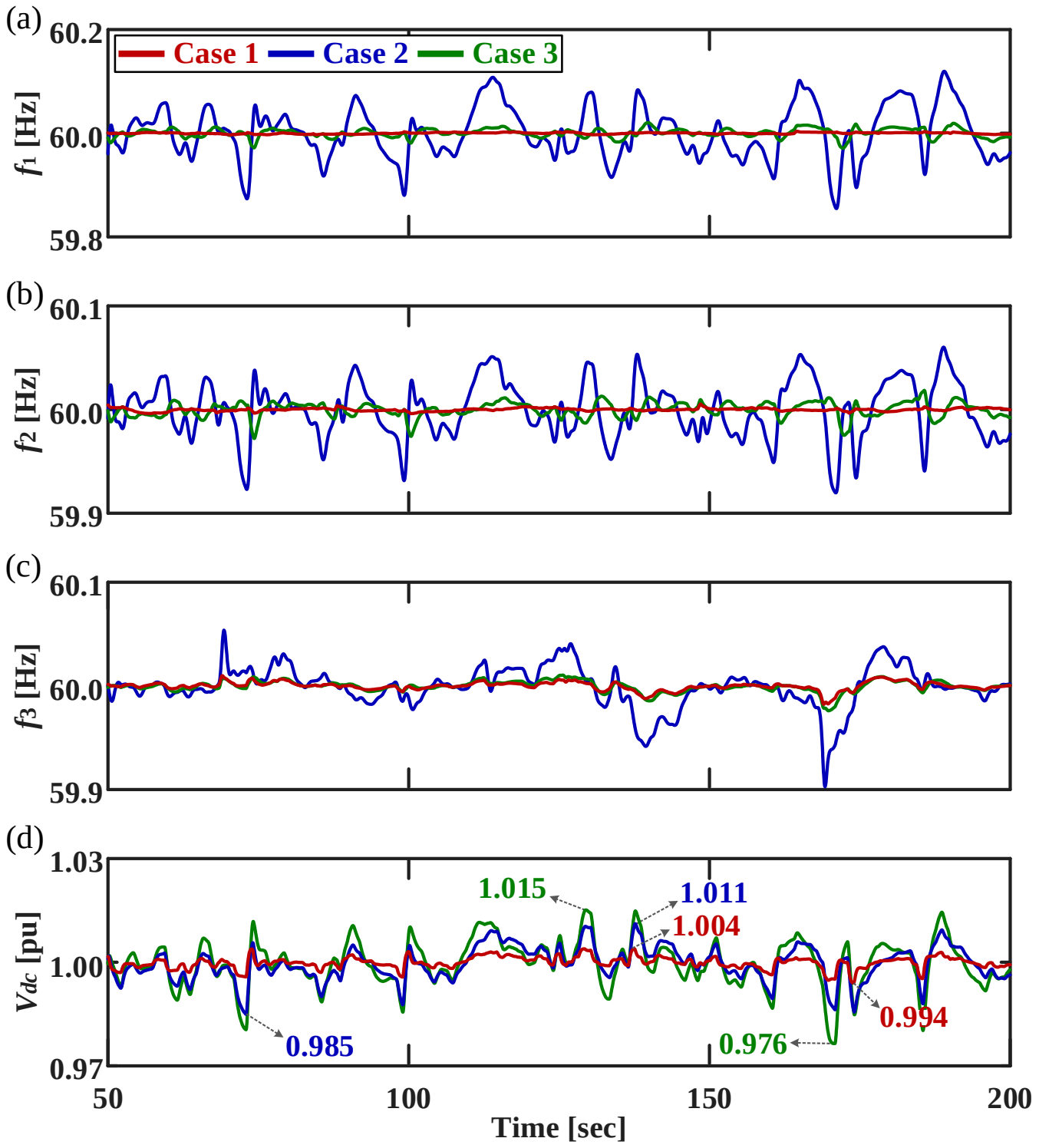

Fig. 10. Responses to the continuous variations in $\Delta P_{L1\text{–}3}$ and $\Delta V_W$ for the proposed and conventional strategies: (a) $f_1$, (b) $f_2$, (c) $f_3$, and (d) $V_{dc}$.

TABLE IV. COMPARISONS OF CONTINUOUS RESPONSE TEST RESULTS

| RMS Variations | Case 1 Individual | Case 1 Total | Case 2 Individual | Case 2 Total | Case 3 Individual | Case 3 Total |
|---|---|---|---|---|---|---|
| $\Delta f_{1,rms}$ [Hz] | 0.002 | | 0.061 | | 0.010 | |
| $\Delta f_{2,rms}$ [Hz] | 0.003 | 0.009 | 0.033 | 0.120 | 0.008 | 0.024 |
| $\Delta f_{3,rms}$ [Hz] | 0.004 | | 0.024 | | 0.006 | |
| $\Delta V_{dc,rms}$ [pu] | 0.002 | | 0.006 | | 0.008 | |

### *E. Performance under Various Conditions*

Additional case studies were conducted to evaluate the performances of the decentralized $H_\infty$ controllers in Cases 1 and 3 under different conditions of the MTDC-linked grids. For example, Fig. 11 represents five different conditions of the inter-grid communication systems: i.e., no failure, single-link and double-link failures, and no active communication. When the communication between grids $k$ and $j$ fails, the decentralized $H_\infty$ controller for grid $k$ generates the optimal reference signals $\mathbf{r}_k$ without receiving the measurements of $\Delta f_j$ and $\Delta V_{dcj}$ from grid $j \neq k$. Fig. 12 shows $f_1$ and $V_{dc}$ for the step responses of the MTDC-linked grids to $\Delta P_{Lk} = 0.1$ pu for all $k$ at $t = 10$ s. In Fig. 12(a), Case 1 enabled the smaller deviation of $f_1$ (and also $f_2$ and $f_3$) and the faster recovery back to the nominal value with almost no overshoot for all the communication conditions than Case 3. This is also the case for variations in $V_{dc}$, as shown in Fig. 12(b). The case study results confirm that the proposed decentralized $H_\infty$ controllers are still more effective and robust in the optimal FR against the events of inter-grid communication failure.

(a)In addition, Fig. 13 compares the step responses of $f_2$ and $f_3$ for variations in $\Delta P_{L1\text{–}3}$ by ±0.1 pu over every 40 s between Cases 1 and 3, when increasing $T_d$ gradually from 20 ms to 50 ms. Note that for brevity, $T_d$ was assumed to be the same for all decentralized $H_\infty$ controllers, as discussed in Section III-C. In Case 3, $T_d$ = 50 ms led to unstable, large oscillations of $f_2$ and $f_3$; this was also the case for $f_1$. It was mainly because in Case 3, the centralized $H_\infty$ control gains were simply divided for the decentralized controllers by setting the off-diagonal elements to zero, as discussed in Section IV-A, which reduced the maximum allowable value of the communication time delay. In other words, the complex-conjugate eigenvalues moved across the imaginary axis from the LHP to the RHP when $T_d$ increased to 50 ms. By contrast, Case 1 still ensured stable, smaller variations in $f_2$ and $f_3$ for all $T_d$ ranging up to 50 ms, although the damped oscillations occurred for large $T_d$. The comparison results prove that the proposed strategy still further enhances the performance and stability of the FR in the MTDC-linked grids when there are relatively large time delays in the communication systems.

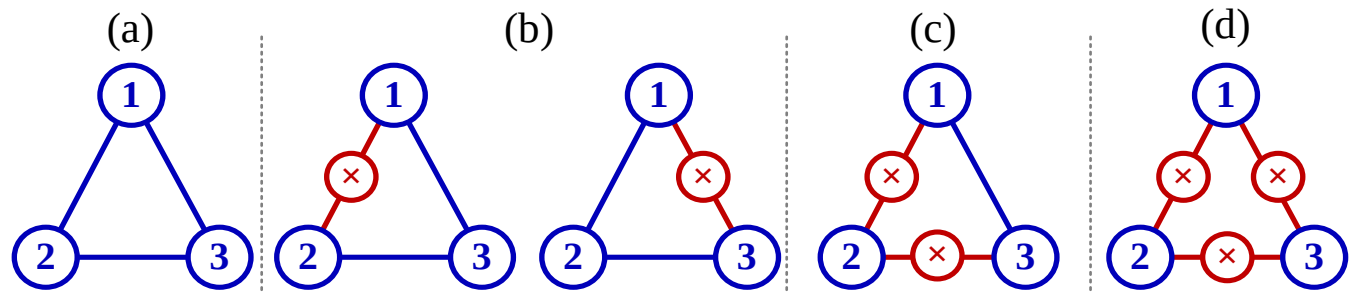

Fig. 11. Failures of the communication links among the MTDC-linked grids: (a) no, (b) single-link, and (c) double-link failures and (d) no communication.

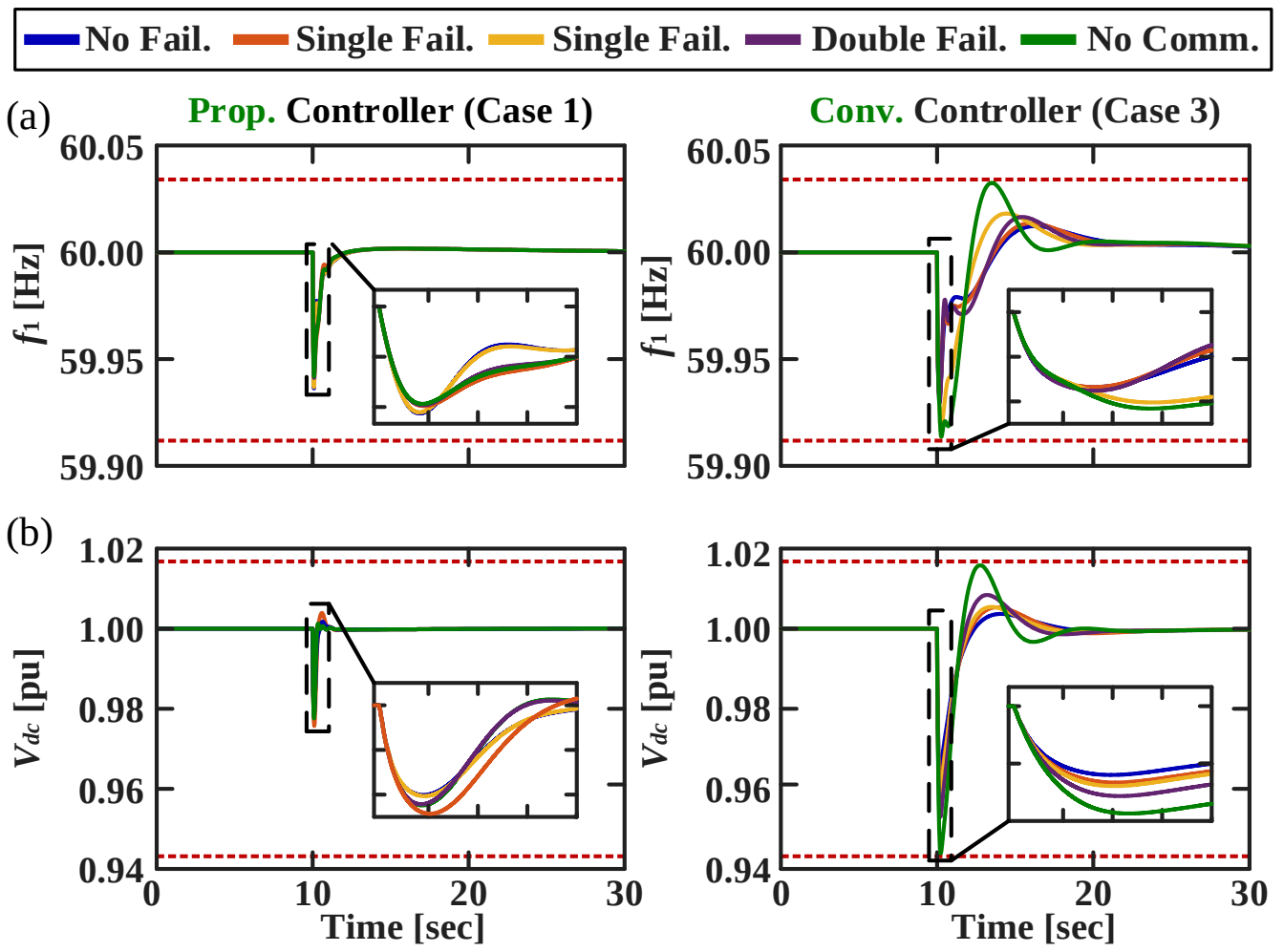

Fig. 12. Step responses of the proposed and conventional $H_\infty$ controllers (i.e., Cases 1 and 3) when the inter-grid communication links fail: (a) $f_1$ and (b) $V_{dc}$.

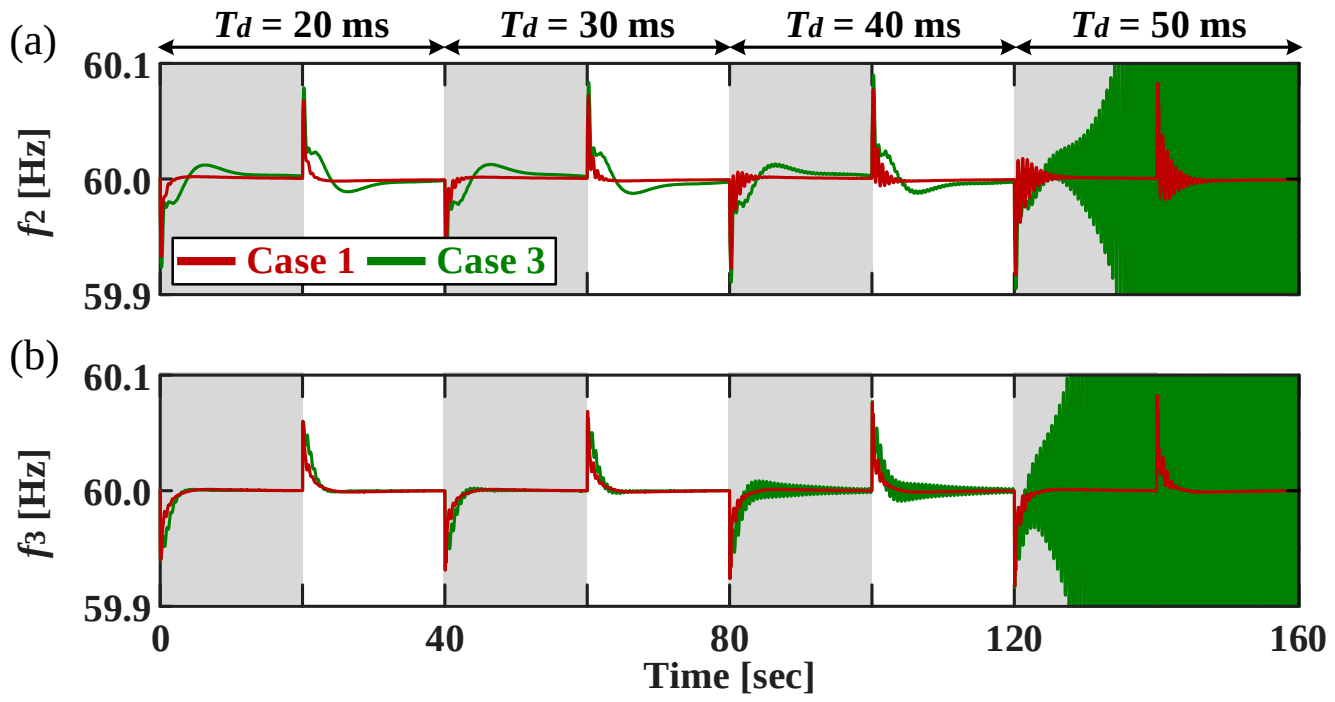

Fig. 13. Comparison between the proposed and conventional $H_\infty$ controllers (i.e., Cases 1 and 3) for $T_d$ ranging from 20 ms to 50 ms: (a) $f_2$ and (b) $f_3$.

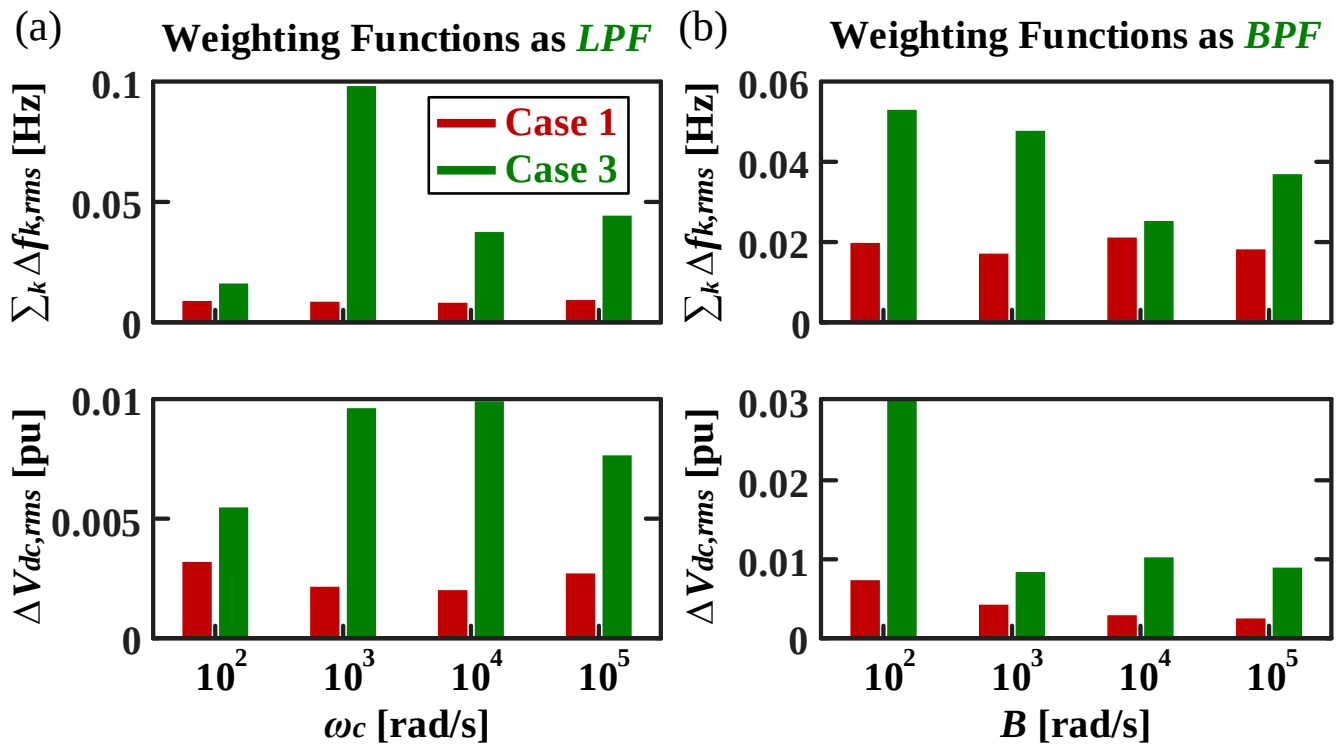

Fig. 14. Comparison between the proposed and conventional $H_\infty$ controllers (i.e., Cases 1 and 3) for different weighting functions: (a) LPF and (b) BPF.

Furthermore, the continuous response tests in Section IV-D were repeated to analyze the performance of the proposed FR strategy with different models of the weighting functions $\mathbf{W}_u$, $\mathbf{W}_e$, and $\mathbf{W}_d$ for $\mathbf{r}_k$, $\mathbf{Y}_{Ek}$, and $\mathbf{d}_k$, respectively (see Fig. 2). Fig. 14(a) shows comparisons between Cases 1 and 3 when the weighting functions were modeled as LPFs with a cutoff frequency $\omega_c$ ranging from $10^2$ rad/s to $10^5$ rad/s, rather than the unity function. Similarly, Fig. 14(b) shows the comparison results when BPFs were adopted with a bandwidth $B$ from $10^2$ rad/s to $10^5$ rad/s. For all different weighting functions, Case 1 led to considerably smaller $\sum_k \Delta f_{k,rms}$ and $\Delta V_{dc,rms}$ than Case 3. This demonstrates the improved effectiveness and robustness of the proposed FR strategy when applied to practical grids with different local controllers of the SGs and MTDC converters and different characteristics of the measurement data and disturbances.

## V. Conclusions

This paper proposes a new strategy for optimal FR in MTDC-linked grids via coordinated, decentralized control of SGs and hybrid converters. A dynamic model of hybrid MTDC-linked grids was implemented considering the decentralized control inputs and outputs. Given the dynamic model, a robust optimization problem was formulated to develop a decentralized $H_\infty$ controller for each grid, considering the inter-grid communication systems and the model parameter uncertainty. A balanced truncation algorithm was applied to reduce the model orders of the decentralized $H_\infty$ controllers, while still preserving their dominant response characteristics. Sensitivity and eigenvalue analyses were then conducted, focusing on the effects of inter-grid measurement, system parameter, uncertainty level, and communication time delay. The results of comparative case studies verified that the proposed FR strategy was more effective in reducing the frequency deviations in all MTDC-linked grids than conventional strategies, while maintaining smaller variations in the dc network voltages. Moreover, compared to the conventional $H_\infty$ control strategy, the proposed strategy also improved the effectiveness, stability, and robustness in the optimal FR of hybrid MTDC-linked grids under various conditions, characterized by inter-grid communication failure, communication time delays, and different weighting functions.

## Appendix

### A. Modeling of AC Grids with Hybrid Converter Interfaces

A regional ac grid includes SGs with reheat steam turbines. For each SG, a synchronous machine is modelled using the 6th-order differential equations, discussed in [31]. A governor and exciter are modeled using the 1st-order transfer functions, based on [32]. The dynamic model of an SG can then be established as:

$$\dot{\mathbf{X}}_{sg} = \mathbf{A}_{sg} \cdot \mathbf{X}_{sg} + \mathbf{B}_{sg} \cdot \mathbf{r}_{sg} + \mathbf{E}_{sg} \cdot \mathbf{I}_{bus}, \quad \text{(A1)}$$

where

$$\mathbf{X}_{sg} = [\Delta E_q', \Delta E_d', \Delta\phi_{1d}, \Delta\phi_{2q}, \Delta\delta_r, \Delta\omega_r, \Delta x_{gov1}, \Delta x_{gov2}, \Delta x_{tur1}, \Delta x_{tur2}, \Delta V_{fd}]^T, \quad \text{(A2)}$$

$$\mathbf{r}_{sg} = [\Delta P_g^{ref}, \Delta E_t^{ref}]^T, \text{ and } \mathbf{I}_{bus} = \Delta I_s^{dq}.$$

Furthermore, in each grid, a phase-locked loop (PLL) is used to measure the grid frequency at the PCC of an MTDC converter and calculate the $dq$ values of the PCC voltages and currents [33]. Considering the PLL operation, an ac transmission network can be modeled as:

$$\dot{\mathbf{X}}_{ac} = \mathbf{A}_{ac} \cdot \mathbf{X}_{ac} + \mathbf{E}_{ac1} \cdot \mathbf{V}_{ac} + \mathbf{E}_{ac2} \cdot \mathbf{I}_{ac} + \mathbf{B}_{acw} \cdot \mathbf{w}_{ac}, \quad \text{(A3)}$$

where $\mathbf{V}_{ac} = [\Delta E_t^d, \Delta E_t^q]^T$, $\mathbf{I}_{ac} = [\Delta I_c^d, \Delta I_c^q]^T$, and $\mathbf{w}_{ac} = \Delta P_L$.

In an MTDC system, a VSC includes inner and outer controller loops with a $P$-$V_{dc}$ droop function. For an LCC, dc terminal voltage or current is controlled to transmit constant or time-varying power from an ac grid to a dc network. The dynamic models of a VSC and an LCC are given [33], [34] as:

$$\dot{\mathbf{X}}_{vsc} = \mathbf{A}_{vsc} \cdot \mathbf{X}_{vsc} + \mathbf{B}_{vsc} \cdot \mathbf{r}_{vsc} + \mathbf{E}_{bus} \cdot \mathbf{V}_{bus} + \mathbf{E}_{dc} \cdot \mathbf{I}_{dc}, \quad \text{(A4)}$$

$$\dot{\mathbf{X}}_{lcc} = \mathbf{A}_{lcc} \cdot \mathbf{X}_{lcc} + \mathbf{B}_{lcc} \cdot \mathbf{r}_{lcc} + \mathbf{E}_{bus} \cdot \mathbf{V}_{bus} + \mathbf{E}_{dc} \cdot \mathbf{I}_{dc}, \quad \text{(A5)}$$

where

$$\mathbf{X}_{vsc} = [\Delta I_c^{dq}, \Delta V_{dc}, \Delta n_{vsc}^{dq}, \Delta m_{vsc}^{dq}]^T, \quad \text{(A6)}$$

$$\mathbf{X}_{lcc} = [\Delta I_c^{dq}, \Delta V_c^{dq}, \Delta I_{dco}, \Delta V_{dc}, \Delta m_{lcc}, \Delta\theta_{lcc}, \Delta n_{lcc}]^T,$$

$$\mathbf{r}_{vsc} = [\Delta P^{ref}, \Delta V_{dc}^{ref}, \Delta V_{mag}^{ref}]^T, \text{ and } \mathbf{r}_{lcc} = \Delta P^{ref}.$$

By combining (A1)–(A6), the dynamic model of ac grid $k$ is represented as:

$$\dot{\mathbf{X}}_k = \mathbf{A}_{k1} \cdot \mathbf{X}_k + \mathbf{B}_k \cdot \mathbf{U}_k + [\mathbf{E}_I \ \mathbf{E}_V] \cdot [\mathbf{I}_k \ \mathbf{V}_k]^T + \mathbf{E}_k \cdot \mathbf{I}_{dc}, \quad \text{(A7)}$$

where

$$\mathbf{I}_k = [\mathbf{I}_{bus}^T \ \mathbf{I}_{ac}^T], \text{ and } \mathbf{V}_k = [\mathbf{V}_{bus}^T \ \mathbf{V}_{ac}^T]. \quad \text{(A8)}$$

Moreover, $[\mathbf{I}_k \ \mathbf{V}_k]^T$ in (A7) can be equivalently expressed as:

$$[\mathbf{I}_k \ \mathbf{V}_k]^T = [\mathbf{F}_I \cdot \mathbf{X}_k \ \mathbf{F}_V \cdot \mathbf{X}_k]^T, \quad \text{(A9)}$$

where $\mathbf{F}_I$ and $\mathbf{F}_V$ are sparse matrices to extract $\mathbf{I}_k$ and $\mathbf{V}_k$ from $\mathbf{X}_k$. Substitution of (A9) into (A7) gives:

$$\dot{\mathbf{X}}_k = \mathbf{A}_k \cdot \mathbf{X}_k + \mathbf{B}_k \cdot \mathbf{U}_k + \mathbf{E}_k \cdot \mathbf{I}_{dc}, \quad \text{(A10)}$$

where $\mathbf{A}_k = \mathbf{A}_{k1} + \mathbf{E}_I \cdot \mathbf{F}_I + \mathbf{E}_V \cdot \mathbf{F}_V$ and $\mathbf{U}_k = [\mathbf{r}_k \ \mathbf{w}_{dk}]^T$. Using (A10), the dynamic model of ac grid $k$ is completed as:

$$\dot{\mathbf{X}}_k = \mathbf{A}_k \cdot \mathbf{X}_k + \mathbf{B}_{rk} \cdot \mathbf{r}_k + \mathbf{B}_{wk} \cdot \mathbf{w}_{dk} + \mathbf{E}_k \cdot \mathbf{I}_{dc}, \quad \text{(A11)}$$

$$\mathbf{Y}_k = \mathbf{C}_k \cdot \mathbf{X}_k, \quad \text{(A12)}$$

where the states, inputs, and outputs are arranged as:

$$\mathbf{X}_k = \begin{cases} [\mathbf{X}_{sg}^T, \ \mathbf{X}_{ac}^T, \ \mathbf{X}_{vsc}^T]^T, & \text{for } k = 1 \text{ and } 2, \\ [\mathbf{X}_{sg}^T, \ \mathbf{X}_{ac}^T, \ \mathbf{X}_{lcc}^T]^T, & \text{for } k = 3, \end{cases} \quad \text{(A13)}$$

$$\mathbf{r}_k = \begin{cases} [\Delta P_{gk}^{ref}, \ \Delta P_k^{ref}, \ \Delta V_{dck}^{ref}]^T, & \text{for } k = 1 \text{ and } 2, \\ [\Delta P_{gk}^{ref}, \ \Delta P_k^{ref}]^T, & \text{for } k = 3, \end{cases} \quad \text{(A14)}$$

$$\mathbf{w}_{dk} = \Delta P_{Lk}, \quad \forall \ k, \quad \text{(A15)}$$

$$\text{and } \mathbf{Y}_k = [\Delta P_{gk}, \ \Delta\omega_{rk}, \ \Delta f_k, \ \Delta V_{magk}, \ \Delta V_{dck}, \ \Delta P_k]^T, \ \forall \ k. \quad \text{(A16)}$$

In (A16), the references for the exciter and PCC voltage variations (i.e., $\Delta E_t^{ref}$ and $\Delta V_{mag}^{ref}$) are set to zero, because this paper focuses on optimal FR.

### *B. Modeling of a PMSG-type OWF and a DC Network*

A PMSG-type OWF is modeled based on [35] and [36], as:

$$\dot{\mathbf{X}}_{owf} = \mathbf{A}_{owf} \cdot \mathbf{X}_{owf} + \mathbf{B}_{owf} \cdot \mathbf{w}_{owf} + \mathbf{E}_{owf} \cdot \mathbf{I}_{dc4}, \quad \text{(B1)}$$

$$\text{where } \mathbf{X}_{owf} = [\Delta\omega_e, \ \Delta\theta_e, \ \Delta I_m^{dq}, \ \Delta V_{dc}, \ \Delta m_{owf}, \ \Delta n_{owf}^{dq}]^T, \quad \text{(B2)}$$

$$\text{and } \quad \mathbf{w}_{owf} = \Delta V_w.$$

In addition, a dc transmission network is characterized using a Π-section model. The corresponding dynamics is linearized as:

$$\Delta \dot{I}_{dc,ij} = (-R_{dc,ij}\Delta I_{dc,ij} + (\Delta V_{dci} - \Delta V_{dcj})) \big/ L_{dc,ij}, \quad \text{(B3)}$$

where $\Delta I_{dc,ij}$ and $\Delta V_{dci}$ are the incremental dc current from bus $i$ to $j$ and the incremental dc voltage at bus $i$, respectively. Moreover, $R_{dc,ij}$ and $L_{dc,ij}$ are the dc resistance and inductance. According to the configuration, the dynamic model of a dc network is given as:

$$\dot{\mathbf{X}}_{dc} = \mathbf{A}_{dc} \cdot \mathbf{X}_{dc} + \mathbf{E}_{dcv} \cdot \mathbf{V}_{dc}, \quad \text{(B4)}$$

$$\text{where } \quad \mathbf{X}_{dc} = [\Delta I_{dc12}, \ \Delta I_{dc13}, \ \Delta I_{dc23}, \ \Delta I_{dc34}]^T, \quad \text{(B5)}$$

$$\text{and } \quad \mathbf{V}_{dc} = [\Delta V_{dc1}, \ \Delta V_{dc2}, \ \Delta V_{dc3}, \ \Delta V_{dc4}]^T.$$